\documentclass[reprint,amsmath,amssymb,aps,prl,superscriptaddress]{revtex4-2}

\usepackage{comment}
\usepackage{enumitem}
\setlist[enumerate]{itemsep=0mm}
\usepackage{color}
\usepackage{graphicx}
\usepackage{hyperref}
\usepackage{braket}
\usepackage{bbm}
\usepackage{tikz}
\usepackage{tikz-3dplot}
\usepackage[outline]{contour}

\begin{document}

\title{
Slow heterogeneous relaxation due to constraints in dual XXZ models}

\author{Lenart Zadnik}
\email{lzadnik@sissa.it}
\affiliation{%
SISSA, Via Bonomea 265, 34136 Trieste, Italy\\
}
\author{Juan P. Garrahan}%
\affiliation{%
School of Physics and Astronomy, University of Nottingham, Nottingham NG7 2RD, United Kingdom
}
\affiliation{%
Centre for the Mathematics and Theoretical Physics of Quantum Non-Equilibrium Systems, University of Nottingham, Nottingham NG7 2RD, United Kingdom
}

\begin{abstract}
With the aim to understand the role of the constraints in the thermalisation of quantum systems, we study the dynamics of a family of kinetically constrained models arising through duality from the XXZ spin chain. We find that integrable and nonintegrable deformations around the stochastic point give rise to ground state phase transitions between localised and delocalised phases, which in turn determine the nature of the relaxation dynamics at finite energy densities. While in the delocalised phase thermalisation is fast and homogeneous, in the localised phase  relaxation is slow, temporal autocorrelations exhibit plateaus indicative of metastability, and the growth of entanglement is heterogeneous in space. Furthermore, by considering relaxation from initial product states, we demostrate that this slow thermalisation can be rationalised directly from the presence of constraints in the dynamics.
\end{abstract}

\maketitle

\noindent
{\bf \em Introduction.} Thermalisation is one of the major challenges to the durability of quantum technologies: quantum coherence---their vital property---cannot be sustained indefinitely due to imperfect isolation from the environment~\cite{brandt1999,cirac2000,zurek2003,schlosshauer2019}. It is also expected to occur in extended isolated systems, where infinitely many degrees of freedom provide an effective bath that leads to equilibration of few-body observables. These attain stationary values predictable by standard statistical ensembles or, in the case of integrable systems with infinitely many conservation laws constraining the dynamics, generalisations thereof~\cite{rigol2007,ilievski2015,gogolin2016,alessio2016}. In generic systems, where only the energy is conserved, one can understand this in the context of the eigenstate thermalisation hypothesis (ETH). The latter, through a combination of thermodynamic suppression of coherences and dephasing, leads to a drastic reduction in the number of parameters required to describe stationarity~\cite{deutsch1991,srednicki1994,rigol2008,deutsch2018}.

While the statistical ensembles can predict the asymptotic expectation values of the few-body observables, they give no information about the time scales over which the relaxation towards them occurs. Speed of relaxation can
be affected by various circumstances, such as the extent to which the symmetries of the physical system are broken by the initial conditions~\cite{alba2017,maric2022,ares2023}, the presence of emergent quasiconserved quantities~\cite{fagotti2014,bertini2015,kemp2017,abanin2017,mori2018}, or dynamical constraints~\cite{horssen2015,lan2018,morningstar2020,pancotti2020,scherg2021,brighi2022,tortora2022,deger2022arresting,deger2022constrained}. The latter are the main feature of kinetically constrained models (KCM), originally conceived as toy models for slow hierarchic dynamics of classical viscous fluids and glasses~\cite{palmer1984,fredrickson1984,ritort2003,garrahan2018aspects}. Mimicking excluded-volume interactions \cite{chandler2010dynamics}---a feature of systems extending from supercooled liquids \cite{berthier2011theoretical,biroli2013perspective} to Rydberg blockade \cite{lesanovsky2011,browaeys2020many-body}---they can lead to a wide variety of exotic quantum nonequilibrium phenomena that have recently been in the spotlight. Examples include jamming and related Hilbert space fragmentation~\cite{yang2020,langlett2021,zadnik2021,pozsgay2021,tamura2022,bidzhiev2022}, quantum many-body scars (QMBS)~\cite{bernien2017,turner2018,moudgalya2022,bluvstein2021}, anomalous transport ~\cite{yang2022,ljubotina2023}, and cooperative dynamics of fractons~\cite{nandkishore2019,pretko2020}. Excluded-volume interactions and dynamical constraints often arise in models with tunable interactions, where strong correlations between excitations are induced in the large coupling limit~\cite{izergin1998,abarenkova2002,bogoliubov2011,lesanovsky2011,pai2020,borla2020,zadnik2021,tartaglia2022,bastianello2022}. Alternatively, KCMs can sometimes be related to such models through duality transformations~\cite{fagotti2022,maric2022,jones2022,eck2023}, an approach we follow here.

\begin{figure}[t!]
    \centering
    \includegraphics[width=0.48\textwidth]{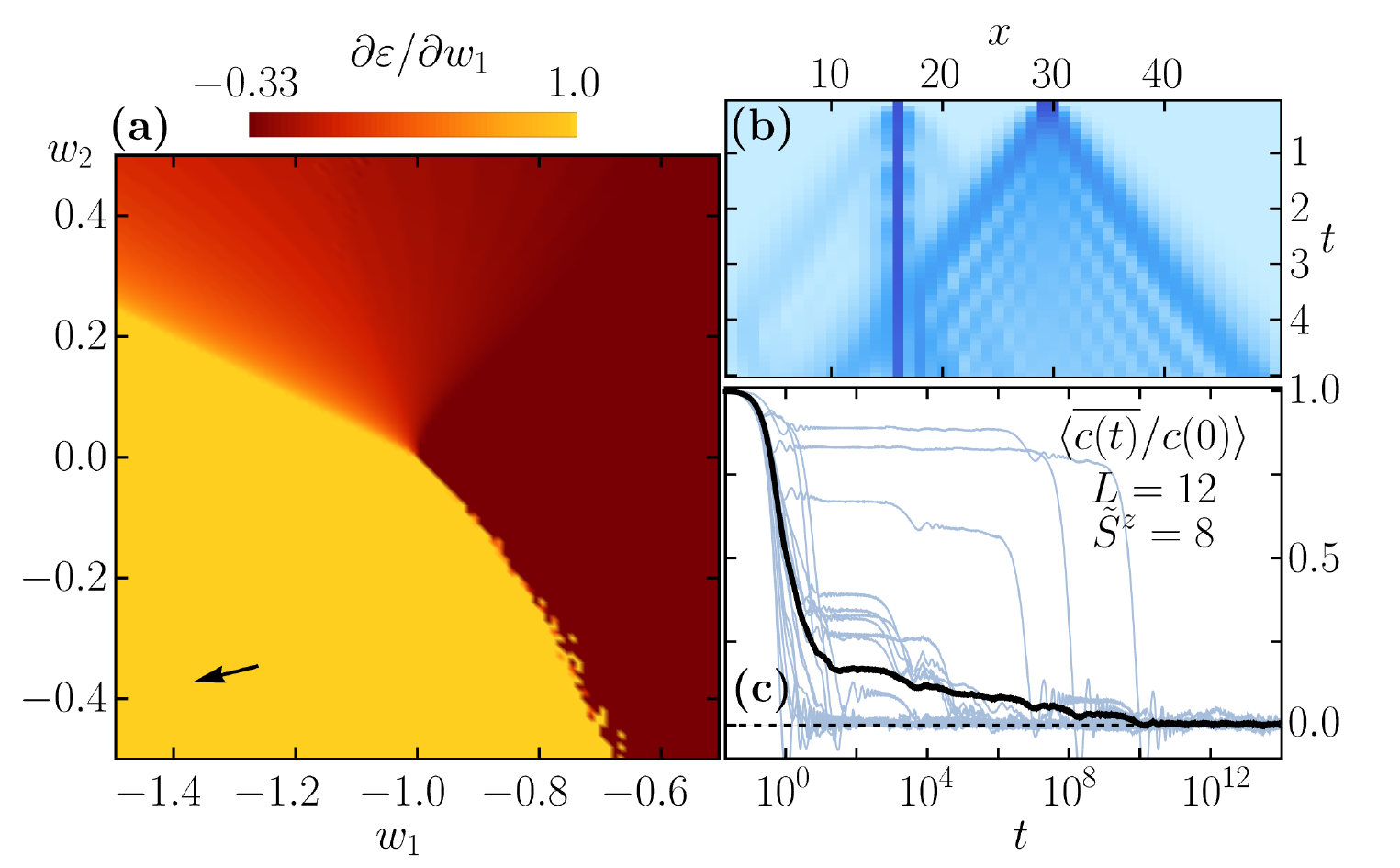}
    \caption{
        {\bf Ground state phases and slow relaxation of the XPX model.} (a) Ground state phase diagram of the XPX model in Eq.~\eqref{eq:Hxpx} for $L=160$ (from DMRG performed using the ITensor library~\cite{itensor,itensor-r0.3}). 
        For smaller $w_{1,2}$
        (yellow region) the ground state is localised, while 
        for larger $w_{1,2}$ (red region) it is delocalised.  
        The arrow indicates the values of $w_{1,2}$ used in panels (b,c). (b) Time evolution of a spin configuration $\ket{\cdots\downarrow\cdots\downarrow\downarrow\cdots}$ with dots denoting spins up (from TEBD based on the Armadillo library~\cite{sanderson2016,sanderson2018}), for $w_1=-2, w_2=-1/2$ in the localised phase. (c) Normalised auto-correlation functions 
        for individual initial configurations
        (light blue) and their average (black) in the localised phase ($w_1=-3.5,w_2=-1.5$), in a specific symmetry sector of the model.
        }
    \label{fig1}
\end{figure}

In this paper we consider a family of one-dimensional (1D) quantum models mappable to the anisotropic Heisenberg spin-$1/2$ chain and its nonintegrable deformation. We investigate two phases of the model depicted in Fig.~\hyperref[fig1]{1(a)}: a phase where the ground state is localised (yellow) and one where it is delocalised (red). A striking feature of the model in the phase with a localised ground state is the emergence of facilitated dynamics at finite energy density, as illustrated in Fig.~\hyperref[fig1]{1(b)}: certain local arrangements (pairs of spins down, see below) can move freely, whereas certain other isolated excitations remain frozen for long times. We demonstrate the resulting separation of time scales by considering the evolution of temporal autocorrelation functions of particle occupation numbers starting from various initial states---cf. Fig.~\hyperref[fig1]{1(c)}.  This in turn gives rise to a 
growth of the bipartite entanglement entropy which is heterogeneous in space, depending on the effect that the dynamical constraints have on spatial fluctuations in initial product states. 

\smallskip

\noindent
{\bf \em Models.} We consider a one-dimensional XPX model on a chain of $L$ spins $1/2$ with open boundary conditions:
\begin{align}
    H_{\rm XPX}\!=\!\sum_{j=2}^{L-1}\!\sigma^x_{j-1}(\mathbbm{1}\!-\!\sigma^z_j)\sigma^x_{j+1}\!+\!w_1 \sigma^z_j\!+\!w_2 \sigma^z_{j-1}\sigma^z_{j+1}.
\label{eq:Hxpx}
\end{align}
Pauli matrices acting in the site $j$ are denoted by $\sigma_j^\alpha$, $\alpha\in\{x,y,z\}$, while  $\mathbbm{1}$ is the identity. The dynamical constraint $\mathbbm{1}\!-\!\sigma_j^z\!=\!2\ket{\downarrow}\!\bra{\downarrow}_j$ allows the spins on sites $j\!-\!1$ and $j\!+\!1$ to flip only if a spin down is between them.

When $w_2=0$ the XPX model is integrable and belongs to a family of models
\begin{align}
\begin{tikzpicture}[baseline=(current  bounding  box.center),scale=1]
\node[anchor=north] at (-2,0) {$H_{\rm XXZ}$};
\draw[->,black,thick] (-2,0) -- (-0.5,1);
\node[anchor=south] at (-0.5,1) {$\hspace{1.5em}H_{\rm XPX}$};
\draw[->,black,thick] (0,1) -- (1.5,0);
\node[anchor=north] at (2,0) {$H_{\rm XOR-FA}$};
\draw[->,black,thick] (1.1,-0.25) -- (-1.4,-0.25);
\node[anchor=north] at (2.22,1.6) {$\sigma_j^z\mapsto\tau^z_{j-1}\tau^z_{j+1}$};
\node[anchor=north] at (2.25,1.1) {$\sigma_{j-1}^x\sigma_{j+1}^x\mapsto\tau^x_j$};
\node[anchor=north] at (-0.2,-0.5)  {$\tau_{j-1}^z\tau_{j}^z\mapsto Z_j$};
\node[anchor=north] at (-0.1,-1)  {$\tau^x_j\mapsto X_jX_{j+1}$};
\node[anchor=north] at (-2.6,1.6)  {$X_j\mapsto \sigma^x_{j-1}\sigma^x_j$};
\node[anchor=north] at (-2.65,1.1)  {$Z_jZ_{j+1}\mapsto\sigma^z_j$};
\draw[fill=green!80!black, opacity=0.1, rounded corners = 2,thick] (0.95,0.5) rectangle ++(2.6,1.1);
\draw[fill=blue!80!green, opacity=0.1, rounded corners = 2,thick] (-3.8,0.5) rectangle ++(2.375,1.1);
\draw[fill=red!80!green, opacity=0.1, rounded corners = 2,thick] (-1.35,-1.6) rectangle ++(2.4,1.1);
\end{tikzpicture}
\label{eq:dual_family}
\end{align}
related by degenerate duality maps often referred to as the bond-site transformations. One of them yields the anisotropic Heisenberg model~\cite{fagotti2022,maric2022,eck2023}
\begin{align}
H_{\rm XXZ}=\sum_{j=2}^{L-1}X_j X_{j+1}+Y_j Y_{j+1}+w_1 Z_j Z_{j+1}
\label{eq:xxz}
\end{align}
and the second one the XOR-Fredrickson-Andersen model
\begin{align}
    H_{\rm XOR\text{-}FA}=\sum_{j=2}^{L-1}\tau_j^x(\mathbbm{1}-\tau_{j-1}^z\tau^z_{j+1})+w_1\tau_{j-1}^z\tau^z_{j+1},
    \label{eq:xor-fa}
\end{align}
whose kinetic constraint---a quantum XOR gate---allows a spin flip to occur only between two oppositely aligned spins~\cite{causer2020}. Operators $X_j,Y_j,Z_j$, and separately $\tau_j^\alpha$, $\alpha\in\{x,y,z\}$, satisfy Pauli algebra and can be represented as Pauli matrices acting in site $j$. For $w_2\ne 0$ the integrability is broken~\cite{maric2022}: the corresponding nonintegrable deformations are $H_{\rm XXZ}+w_2\, Z_{j-1}Z_j Z_{j+1}Z_{j+2}$ and $H_{\rm XOR\text{-}FA}+w_2\, \tau^z_{j-2}\tau^z_{j+2}$.

We note that there is some freedom in specifying the duality transformations. Choosing  $X_j\!\mapsto\! \sigma^x_{j-1}\sigma^x_{j}$, $Y_j\!\mapsto\!\sigma^x_{j-1}\sigma^y_j\sigma^z_{j+1}\cdots\sigma^z_L$, $Z_j\!\mapsto\!\sigma^z_j\cdots\sigma^z_L$ for $1\le j\le L$, with convention $\sigma_0^x=1$, the conserved magnetisation $S^z=\sum_{j=1}^L Z_j$ of the Heisenberg model is mapped into the ``semilocal'' charge $\tilde{S}^z=\sum_{j=1}^L\sigma_j^z\cdots\sigma^z_L$ of the XPX model: $[H_{\rm XPX},\tilde{S}^z]=0$. Despite not being local, such an operator may crucially affect local relaxation~\cite{fagotti2022,maric2022}.

\begin{figure}[t!]
    \centering
    \includegraphics[width=0.48\textwidth]{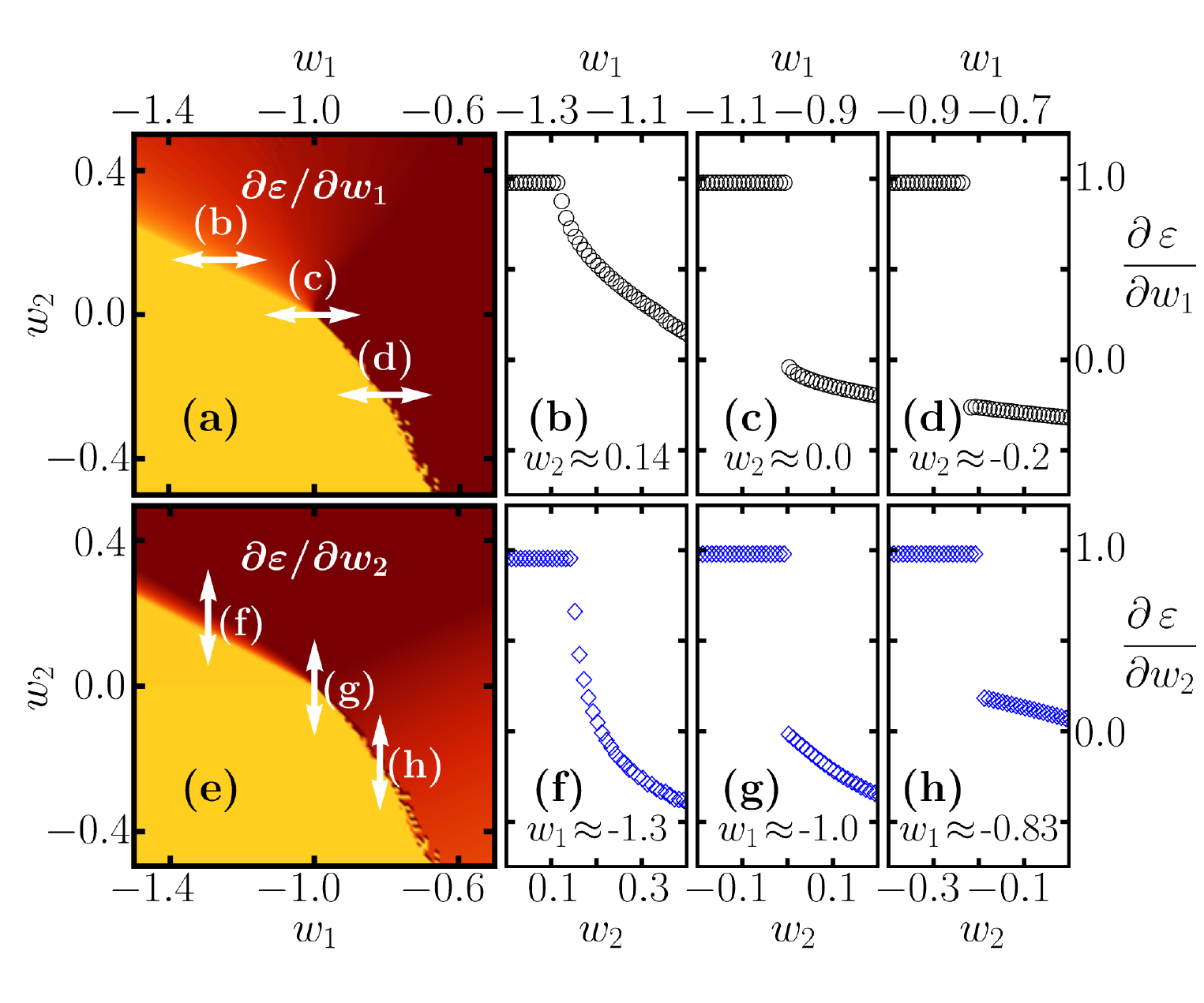}
    \caption{
        {\bf Order of ground state phase transitions in the XPX model.} 
        Panels in the first and the second row show the derivatives of the ground state energy on $w_1$, resp. $w_2$. For $w_2\le 0$ the transition between the localised and the delocalised phase (i.e., between the regions color-coded yellow and red, respectively) is a first-order transition. For $w_2>0$ it is of the second order.
        }
    \label{fig2}
\end{figure}

Notably, all of the models in the family~\eqref{eq:dual_family} have classical stochastic counterparts. The one for the XPX model with $w_2=0$ and $w_1=-1-s$ is associated to $\mathbbm{W}(s)=-U(H_{\rm XPX}+(1+s)\mathbbm{1})U^{-1}$,
where $U\!=\!\prod_{j=1}^{L/4}\!\sigma^z_{4j-1}\sigma^z_{4j}$ and we have assumed $L/4 \in \mathbbm{N}$ for convenience. 
The operator $\mathbbm{W}(s=0)$ is a stochastic Markov generator, while for $s \neq 0$ it is a deformed (or ``tilted'') generator encoding the large deviation (LD) statistics~\cite{touchette2009,garrahan2018aspects,jack2020ergodicity} of the number of spin-flips (dynamical activity~\cite{lecomte2005,lecomte2007,garrahan2009}) in trajectories of the dynamics. 

Via duality to the XXZ model, and up to trivial boundaries, $\mathbbm{W}(s=0)$ 
corresponds to the stochastic generator of the 
classical {\em symmetric simple exclusion process} (SSEP)~\cite{golinelli2006}, and for $s \neq 0$, it encodes the LDs of the activity in the SSEP 
\cite{appert-rolland2008universal,jack2015hyperuniformity}. The SSEP is known to have a phase transition in the space of its (long-time) stochastic trajectories between an active and an inactive phase, which shows up as a nonanalyticity at $s=0$ in the largest eigenvalue of 
$\mathbbm{W}(s)$ (in the large size limit) \cite{appert-rolland2008universal,lecomte2012inactive,jack2015hyperuniformity}. 
The dynamical LD method \cite{touchette2009,garrahan2018aspects,jack2020ergodicity} provides a means for a statistical ensemble description of trajectories, and the ensuing dynamical phase transition, in the classical stochastic SSEP \cite{appert-rolland2008universal,lecomte2012inactive,jack2015hyperuniformity}. In the dual picture for the quantum model, this transition (occurring at $w_1=-1$) corresponds to the ferromagnetic-paraferromagnetic phase transition in the ground state of the XXZ model~\cite{cloizeaux1966,yang1966}. In what follows we explore how it affects the relaxation in the XPX model.

\smallskip

\noindent
{\bf \em Localised and delocalised phases.} In the context of quantum dynamics the inactive and active phase of the XPX model will be referred to as the localised and delocalised phase, respectively. Choosing the inverse participation ratio ${\rm IPR}=\sum_{\psi} |\braket{\psi|{\rm GS}}|^4$ as a measure of localisation ($\ket{\rm GS}$ is the ground state and $\ket{\psi}$ are computational basis states), we indeed find ${\rm IPR}$ close to one in the former and close to zero in the latter. As shown in Fig.~\ref{fig2} [see also Fig.~\hyperref[fig1]{1(a)}], these two phases extend beyond the integrable line $w_2=0$. For all $w_2\le0$ they are separated by a first order transition, both along $w_1$ as well as $w_2$. Instead, for $w_2>0$ the transition is a second order one, cf.\ Fig.~\ref{fig2}. 

An interesting feature of the localised phase, indicated in Fig.~\hyperref[fig1]{1(b)}, is what could be described as ``fractonic'' nature of the excitations~\cite{pai2020}: an isolated spin down remains immobile for long times, while two adjecent spins down can move without energy costs. We note that the isolated spin down in the integrable XPX model ($w_2=0$) corresponds to a domain wall in the XXZ model, which does not melt in the $|w_1|>1$ regime due to being close to a stable kink solution~\cite{koma1998,gobert2005,mossel2010,misguich2017,gamayun2019,misguich2019}.
The fractonic dynamics in which particles can move only if paired (assisted hopping) is a sort of dynamical facilitation~\cite{ritort2003}, which can lead to separation of time scales~\cite{garrahan2018aspects}. Remarkably, the resulting metastability exhibited by correlation functions which involve the entire spectrum of $H_{\rm XPX}$, and which will be explored in the following, is tied to the localisation of the ground state.

\begin{figure}[t!]
    \centering
    \includegraphics[width=0.48\textwidth]{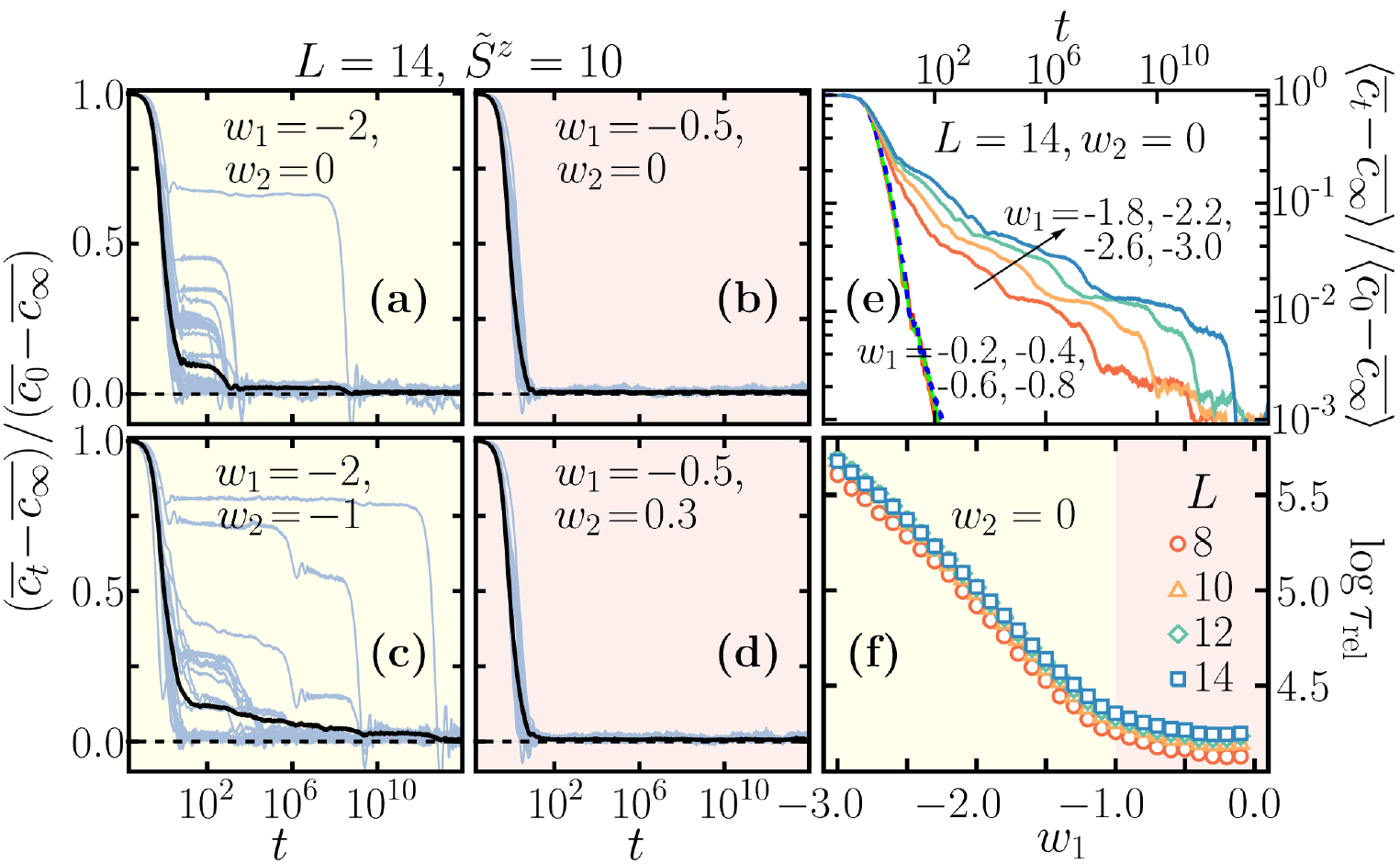}
    \caption{
        {\bf Slowdown of the relaxation in the localised phase.} 
        (a-d) Normalised time-integrated autocorrelations $(\overline{c_t}-\overline{c_\infty})/(\overline{c_0}-\overline{c_\infty})$ from initial computational basis states (light blue), and their average over the sector $\tilde{S}^z=L-4$ (black), for $L=14$. Panels (a,b) are for the integrable case ($w_2=0$), panels (c,d) for the non-integrable one ($w_2 \neq 0$). Panels (a,c) are in the localised phase, panels (b,d) in the delocalised one. 
        (e) Normalised time-integrated correlation $\braket{\overline{c_t}-\overline{c_\infty}}$ averaged over all computational basis states (all sectors of Hilbert space). In the delocalised regime (overlapping curves, topmost being dashed) there is almost no $w_1$ dependence of the relaxation time, while in the localised regime there is a clear slowdown of relaxation with increasing $|w_1|$. The relaxation time seems to obey exponential scaling $\tau_{\rm rel}\sim \exp(\alpha|w_1+1|)$.
        (f) Estimate of $\log\tau_{\rm rel}$ from the area under $\braket{\overline{c_t}-\overline{c_\infty}}/\braket{\overline{c_0}-\overline{c_\infty}}$ as a function of $\log t$, averaged over the entire Hilbert space.
        }
    \label{fig3}
\end{figure}

\begin{figure*}[t!]
    \centering
    \includegraphics[width=\textwidth]{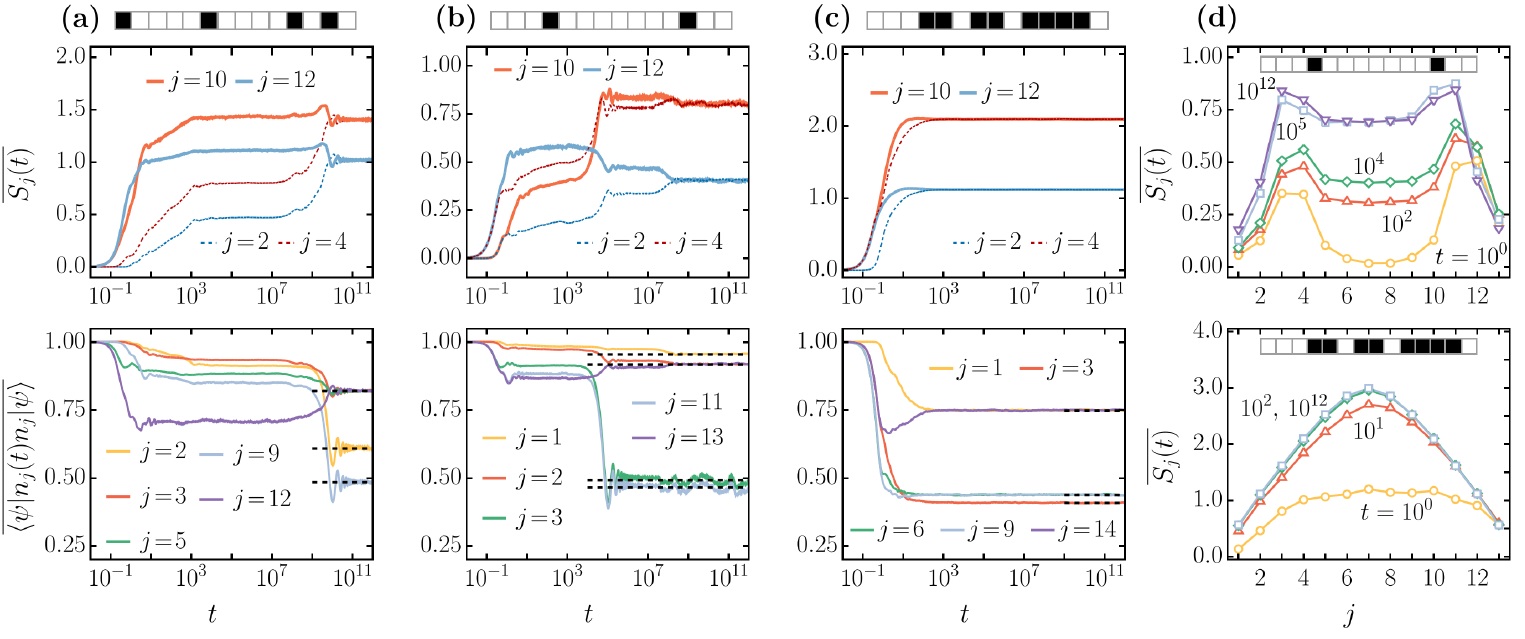}
    \caption{
        {\bf Dynamic heterogeneity of the entanglement entropy.} 
        (a-c) Evolution of the time-integrated bipartite entanglement $\overline{S_j(t)}$ (top panels), and the time-integrated temporal autocorrelation of the one-site occupation number $\overline{\bra{\psi}n_j(t)n_j\ket{\psi}}$ (bottom panels), 
        starting from three different initial computational basis states $\ket{\psi}$ (black squares denote spins down, white ones spins up). The plots are for the localised regime of the nonintegrable model with $w_1=-2,w_2=-1/2$. 
        (d) Snapshots of the EE profile at different times. The initial configuration for which the time autocorrelations have plateaus indicative of slow relaxation exhibits spatially heterogeneous entanglement evolution. The evolution of EE in the configuration with fast relaxation is instead faster and homogeneous.
        }
    \label{fig4}
\end{figure*}

\smallskip

\noindent
{\bf \em Slow relaxation.} To probe metastability we consider the average temporal autocorrelation of the one-site occupation number $n_j=(\mathbbm{1}+\sigma_j^z)/2$:
\begin{align}
   c_t=\frac{1}{L}\sum_{j=1}^L \bra{\psi}   n_j(t)n_j\ket{\psi}.
   \label{eq:correlator}
\end{align}
Here, $n_j(t)=e^{i H_{\rm XPX} t}n_je^{-i H_{\rm XPX} t}$ and $\ket{\psi}$
is a computational basis product state (an eigenstate of all $\sigma_j^z$). For such initial states $c_t$ corresponds to the average magnetisation of the initially occupied sites at time $t$. To smooth out fast fluctuations we furthermore define $\overline{c_t}=t^{-1}\int_0^t{\rm d}\tau\,c(\tau)$, which asymptotically approaches the diagonal-ensemble prediction $\overline{c_\infty}=L^{-1}\sum_j\sum_{E_\ell=E_m}\psi_\ell^*\psi^{}_m \bra{\ell}n_j\ket{m}$, the sum over $j$ running over the initially occupied sites only.

Panels (a-d) in Fig.~\ref{fig3} portray the autocorrelation functions $(\overline{c_t}-\overline{c_\infty})/(\overline{c_0}-\overline{c_\infty})$, normalised so as to lie between $0$ and $1$, for a selection of initial states $\ket{\psi}$ in the semilocal-charge sector $\tilde{S}^z=L-4$ with $L=14$. Plotted is also the infinite-temperature average $\braket{(\overline{c_t}-\overline{c_\infty})/(\overline{c_0}-\overline{c_\infty})}$ over all $\ket{\psi}$ in that sector.
In contrast to the delocalised phase, panels (b,d) (red background), where all correlation functions quickly attain stationary values, a large number of correlators in the localised phase, panels (a,c) (yellow background), exhibit plateaus which persist for long times before finally relaxing. 

In the nonintegrable case some initial states relax in stages, see Fig.~\hyperref[fig3]{3(c)}. The resulting hierarchical decay of the average correlator is typical for classical glassy systems, where it is associated with a sequence of different length scales on which relaxation occurs~\cite{sollich1999,ashton2005,keys2011}. Instead, in the integrable case hierarchical relaxation is noticeable in the autocorrelation $\braket{\overline{c_t}-\overline{c_\infty}}/\braket{\overline{c_0}-\overline{c_\infty}}$, where the averages are taken over the entire Hilbert space, see Fig.~\hyperref[fig3]{3(e)}. Defining the relaxation time $\tau_{\rm rel}$ as the one required by the average correlator to fall below a certain cutoff value $\varepsilon$, there is a clear distinction between the delocalised phase, with no dependence on $w_1$, and the localised one, for which Fig.~\hyperref[fig3]{3(e)} suggests $\tau_{\rm rel}\sim e^{\alpha |w_1+1|}$ for some $\alpha>0$ which may differ between the successive plateaus. 

An alternative estimate for the relaxation time scale is the area under the averaged correlator in the logarithmic time scale: $\log\tau_{\rm rel}\approx \int_{\log t_{\rm min}}^\infty{\rm d}[\log t] (\braket{\overline{c_t}-\overline{c_\infty}}/\braket{\overline{c_0}-\overline{c_\infty}})$
\footnote{We choose the short-time cutoff $t_{\rm min}=10^{-2}$, but this value does not affect the estimation of the relaxation time.}. 
We show this $\tau_{\rm rel}$ in Fig.~\hyperref[fig3]{3(f)} as a function of $w_1$ ranging between the delocalised and the localised regime of the integrable model ($w_2=0$): there is a clear crossover from a regime where $\tau_{\rm rel}$ is only weakly dependent on $w_1$, coinciding with the delocalised phase (red background), to one of exponential dependence on $w_1$ in the localised one (yellow background). 
Note the lack of dependence on system size for the sizes 
accessible to our numerics. This indicates that relaxation can be slow but not divergent with system size, a typical feature of glassy dynamics.

Note that through duality between the models, cf.\ Eq.~\eqref{eq:dual_family}, the metastability presented above should occur also for the autocorrelation function of the domain-wall occupation number $(\mathbbm{1}-Z_j Z_{j+1})/2$ in the XXZ model, as well as for the autocorrelation function of $(\mathbbm{1}-\tau^z_{j-1}\tau^z_{j+1})/2$ in the XOR-FA model.

\smallskip

\noindent
{\bf \em Large coupling regime.} Slow relaxation observed above should be contrasted with the one in the strong coupling limit of the XPX model. In particular, for $w_2=0$ and $w_1\to-\infty$ the dynamics of the XPX model is described by the integrable {\em dual folded XXZ model}~\cite{zadnik2021,zadnik2021hydro,pozsgay2021}. The latter has an exponentially large sector of jammed states, typical for Rydberg blockade systems~\cite{lesanovsky2012,bernien2017}, and exhibits strong Hilbert space fragmentation~\cite{yang2020,bidzhiev2022,bocini2022}. 
For finite $w_1$ the dual folded XXZ model accurately describes the time evolution of the XPX model up to times $t\sim |w_1|$. On such time scales the dynamics is confined to small subsectors of Hilbert space and time-averaged correlation functions $\overline{c_t}$ exhibit plateaus. While we have checked that they can be correctly predicted by the folded model's diagonal ensemble, such plateaus are not observed for the values of $w_{1,2}$ considered herein (see Ref.~\cite{michailidis2018} for a perturbative picture of prerelaxation in certain nonintegrable deformations of the XXZ model).

\smallskip

\noindent
{\bf \em Dynamic heterogeneity in entanglement.} Entropy growth provides crucial insight into the role of kinetic constraints in the emergence of metastability~\cite{garrahan2002,merolle2005,lan2018,horssen2015}. To demonstrate the heterogeneity of dynamical facilitation we consider the bipartite entanglement entropy (EE) $S_j(t)=-{\rm Tr}[\rho_{j}(\tau)\log\rho_{j}(\tau)]$, where $\rho_j(t)$ is the time-evolved reduced density matrix of the subsystem consisting of sites $1,2,\ldots j$. Panels (a-c) of Fig.~\ref{fig4} show $\overline{S_j(t)}=t^{-1}\int_0^t{\rm d}\tau S_j(\tau)$ for several initial states in the localised phase ($w_1=-2$, $w_2=-1/2$) of the XPX model on $L=14$ sites. Figure~\hyperref[fig4]{4(a)} shows the difference between the EE growth when spins down which facilitate relaxation are initially closer (faster EE growth) to when they are further apart (slower EE growth). Note that the heterogeneity of the EE evolution is accompanied by plateaus in the correlators $\overline{\bra{\psi}n_j(t)n_j\ket{\psi}}$. Figure~\hyperref[fig4]{4(b)} illustrates
the interplay of the kinetic term and the three-site potential energy term (for $w_2\ne 0$): facilitation caused by the spin down closer to the boundary affects less neighbouring sites, which are thus entangled faster by the quantum unitary dynamics, that is, $\overline{S_{12}(t)}\ge\overline{S_{2}(t)}$ for all $t$. Finally, due to the assisted hopping in the localised regime, a large density of paired down spins results in a quick equilibration, as shown in Fig.~\hyperref[fig4]{4(c)}. The profiles of EE plotted at different times in Fig.~\hyperref[fig4]{4(d)} further corroborate the observation that metastability is associated with dynamic heterogeneity, as is also the case in classical glassy materials with or without quenched disorder~\cite{berthier2011theoretical,berthier2011,garrahan2002,chandler2010dynamics}.

\smallskip

\noindent
{\bf \em Discussion.} We have investigated how metastability and slow heterogeneous relaxation emerge from the kinetic constraints in the XPX spin chain (and by extension in its duals, the XXZ and XOR-FA models). The onset of anomalously slow dynamics coincides with a ground state phase transition from a delocalised to a localised one. This is similar to what occurs in other 1D and 2D constrained models \cite{horssen2015,lan2018,pancotti2020,tortora2022} for deformations around their stochastic (frustration free) points. 
In our case, we also find that the two phases with distinct relaxation extend beyond the range of parameters for which the model is integrable and which include the stochastic point. Interestingly, another contrast to previous results is that the ground state transitions delimiting the two dynamical regimes are not always first-order. 

While the models studied herein bear some resemblance to certain KCMs with a variety of low-entangled nonthermalising eigenstates~\cite{pancotti2020,iadecola2020}, the methods for constructing such states do not straightforwardly generalise here due to crucial differences in either dynamical facilitation, or interaction. Whether the onset of slow heterogeneous dynamics in quantum KCMs is related to nonthermalising states interwoven into the energy spectrum, or the presence of some other exotic symmetries constraining the dynamics~\cite{moudgalya2022,buca2023,borsi2023} remains one of the intriguing open questions.

\smallskip

\noindent
{\bf \em Acknowledgements.}
LZ thanks Enej Ilievski for valuable discussions. This work was supported by the ERC under Consolidator grant no. 771536 NEMO (LZ) and Starting grant no. 805252 LoCoMacro (LZ), and by the EPSRC  under Grant no. EP/V031201/1 (JPG).

\bibliography{references.bib}

\begin{thebibliography}{98}%
\makeatletter
\providecommand \@ifxundefined [1]{%
 \@ifx{#1\undefined}
}%
\providecommand \@ifnum [1]{%
 \ifnum #1\expandafter \@firstoftwo
 \else \expandafter \@secondoftwo
 \fi
}%
\providecommand \@ifx [1]{%
 \ifx #1\expandafter \@firstoftwo
 \else \expandafter \@secondoftwo
 \fi
}%
\providecommand \natexlab [1]{#1}%
\providecommand \enquote  [1]{``#1''}%
\providecommand \bibnamefont  [1]{#1}%
\providecommand \bibfnamefont [1]{#1}%
\providecommand \citenamefont [1]{#1}%
\providecommand \href@noop [0]{\@secondoftwo}%
\providecommand \href [0]{\begingroup \@sanitize@url \@href}%
\providecommand \@href[1]{\@@startlink{#1}\@@href}%
\providecommand \@@href[1]{\endgroup#1\@@endlink}%
\providecommand \@sanitize@url [0]{\catcode `\\12\catcode `\$12\catcode
  `\&12\catcode `\#12\catcode `\^12\catcode `\_12\catcode `\%12\relax}%
\providecommand \@@startlink[1]{}%
\providecommand \@@endlink[0]{}%
\providecommand \url  [0]{\begingroup\@sanitize@url \@url }%
\providecommand \@url [1]{\endgroup\@href {#1}{\urlprefix }}%
\providecommand \urlprefix  [0]{URL }%
\providecommand \Eprint [0]{\href }%
\providecommand \doibase [0]{https://doi.org/}%
\providecommand \selectlanguage [0]{\@gobble}%
\providecommand \bibinfo  [0]{\@secondoftwo}%
\providecommand \bibfield  [0]{\@secondoftwo}%
\providecommand \translation [1]{[#1]}%
\providecommand \BibitemOpen [0]{}%
\providecommand \bibitemStop [0]{}%
\providecommand \bibitemNoStop [0]{.\EOS\space}%
\providecommand \EOS [0]{\spacefactor3000\relax}%
\providecommand \BibitemShut  [1]{\csname bibitem#1\endcsname}%
\let\auto@bib@innerbib\@empty
\bibitem [{\citenamefont {Brandt}(1999)}]{brandt1999}%
  \BibitemOpen
  \bibfield  {author} {\bibinfo {author} {\bibfnamefont {H.~E.}\ \bibnamefont
  {Brandt}},\ }\bibfield  {title} {\bibinfo {title} {{Qubit devices and the
  issue of quantum decoherence}},\ }\href
  {https://www.sciencedirect.com/science/article/pii/S0079672799000038}
  {\bibfield  {journal} {\bibinfo  {journal} {Prog. Quantum Electron.}\
  }\textbf {\bibinfo {volume} {22}},\ \bibinfo {pages} {257} (\bibinfo {year}
  {1999})}\BibitemShut {NoStop}%
\bibitem [{\citenamefont {Cirac}\ and\ \citenamefont
  {Zoller}(2000)}]{cirac2000}%
  \BibitemOpen
  \bibfield  {author} {\bibinfo {author} {\bibfnamefont {J.~I.}\ \bibnamefont
  {Cirac}}\ and\ \bibinfo {author} {\bibfnamefont {P.}~\bibnamefont {Zoller}},\
  }\bibfield  {title} {\bibinfo {title} {{A scalable quantum computer with ions
  in an array of microtraps}},\ }\href {https://doi.org/10.1038/35007021}
  {\bibfield  {journal} {\bibinfo  {journal} {Nature}\ }\textbf {\bibinfo
  {volume} {404}},\ \bibinfo {pages} {579} (\bibinfo {year}
  {2000})}\BibitemShut {NoStop}%
\bibitem [{\citenamefont {Zurek}(2003)}]{zurek2003}%
  \BibitemOpen
  \bibfield  {author} {\bibinfo {author} {\bibfnamefont {W.~H.}\ \bibnamefont
  {Zurek}},\ }\bibfield  {title} {\bibinfo {title} {{Decoherence, einselection,
  and the quantum origins of the classical}},\ }\href
  {https://doi.org/10.1103/RevModPhys.75.715} {\bibfield  {journal} {\bibinfo
  {journal} {Rev. Mod. Phys.}\ }\textbf {\bibinfo {volume} {75}},\ \bibinfo
  {pages} {715} (\bibinfo {year} {2003})}\BibitemShut {NoStop}%
\bibitem [{\citenamefont {Schlosshauer}(2019)}]{schlosshauer2019}%
  \BibitemOpen
  \bibfield  {author} {\bibinfo {author} {\bibfnamefont {M.}~\bibnamefont
  {Schlosshauer}},\ }\bibfield  {title} {\bibinfo {title} {{Quantum
  decoherence}},\ }\href
  {https://www.sciencedirect.com/science/article/pii/S0370157319303084}
  {\bibfield  {journal} {\bibinfo  {journal} {Phys. Rep.}\ }\textbf {\bibinfo
  {volume} {831}},\ \bibinfo {pages} {1} (\bibinfo {year} {2019})}\BibitemShut
  {NoStop}%
\bibitem [{\citenamefont {Rigol}\ \emph {et~al.}(2007)\citenamefont {Rigol},
  \citenamefont {Dunjko}, \citenamefont {Yurovsky},\ and\ \citenamefont
  {Olshanii}}]{rigol2007}%
  \BibitemOpen
  \bibfield  {author} {\bibinfo {author} {\bibfnamefont {M.}~\bibnamefont
  {Rigol}}, \bibinfo {author} {\bibfnamefont {V.}~\bibnamefont {Dunjko}},
  \bibinfo {author} {\bibfnamefont {V.}~\bibnamefont {Yurovsky}},\ and\
  \bibinfo {author} {\bibfnamefont {M.}~\bibnamefont {Olshanii}},\ }\bibfield
  {title} {\bibinfo {title} {{Relaxation in a completely integrable many-body
  quantum system: An ab initio study of the dynamics of the highly excited
  states of 1D lattice hard-core bosons}},\ }\href
  {https://doi.org/10.1103/PhysRevLett.98.050405} {\bibfield  {journal}
  {\bibinfo  {journal} {Phys. Rev. Lett.}\ }\textbf {\bibinfo {volume} {98}},\
  \bibinfo {pages} {050405} (\bibinfo {year} {2007})}\BibitemShut {NoStop}%
\bibitem [{\citenamefont {Ilievski}\ \emph {et~al.}(2015)\citenamefont
  {Ilievski}, \citenamefont {De~Nardis}, \citenamefont {Wouters}, \citenamefont
  {Caux}, \citenamefont {Essler},\ and\ \citenamefont {Prosen}}]{ilievski2015}%
  \BibitemOpen
  \bibfield  {author} {\bibinfo {author} {\bibfnamefont {E.}~\bibnamefont
  {Ilievski}}, \bibinfo {author} {\bibfnamefont {J.}~\bibnamefont {De~Nardis}},
  \bibinfo {author} {\bibfnamefont {B.}~\bibnamefont {Wouters}}, \bibinfo
  {author} {\bibfnamefont {J.-S.}\ \bibnamefont {Caux}}, \bibinfo {author}
  {\bibfnamefont {F.~H.~L.}\ \bibnamefont {Essler}},\ and\ \bibinfo {author}
  {\bibfnamefont {T.}~\bibnamefont {Prosen}},\ }\bibfield  {title} {\bibinfo
  {title} {{Complete Generalized Gibbs Ensembles in an Interacting Theory}},\
  }\href {https://doi.org/10.1103/PhysRevLett.115.157201} {\bibfield  {journal}
  {\bibinfo  {journal} {Phys. Rev. Lett.}\ }\textbf {\bibinfo {volume} {115}},\
  \bibinfo {pages} {157201} (\bibinfo {year} {2015})}\BibitemShut {NoStop}%
\bibitem [{\citenamefont {Gogolin}\ and\ \citenamefont
  {Eisert}(2016)}]{gogolin2016}%
  \BibitemOpen
  \bibfield  {author} {\bibinfo {author} {\bibfnamefont {C.}~\bibnamefont
  {Gogolin}}\ and\ \bibinfo {author} {\bibfnamefont {J.}~\bibnamefont
  {Eisert}},\ }\bibfield  {title} {\bibinfo {title} {{Equilibration,
  thermalisation, and the emergence of statistical mechanics in closed quantum
  systems}},\ }\href {https://dx.doi.org/10.1088/0034-4885/79/5/056001}
  {\bibfield  {journal} {\bibinfo  {journal} {Rep. Prog. Phys.}\ }\textbf
  {\bibinfo {volume} {79}},\ \bibinfo {pages} {056001} (\bibinfo {year}
  {2016})}\BibitemShut {NoStop}%
\bibitem [{\citenamefont {D'Alessio}\ \emph {et~al.}(2016)\citenamefont
  {D'Alessio}, \citenamefont {Kafri}, \citenamefont {Polkovnikov},\ and\
  \citenamefont {Rigol}}]{alessio2016}%
  \BibitemOpen
  \bibfield  {author} {\bibinfo {author} {\bibfnamefont {L.}~\bibnamefont
  {D'Alessio}}, \bibinfo {author} {\bibfnamefont {Y.}~\bibnamefont {Kafri}},
  \bibinfo {author} {\bibfnamefont {A.}~\bibnamefont {Polkovnikov}},\ and\
  \bibinfo {author} {\bibfnamefont {M.}~\bibnamefont {Rigol}},\ }\bibfield
  {title} {\bibinfo {title} {{From quantum chaos and eigenstate thermalization
  to statistical mechanics and thermodynamics}},\ }\href
  {https://doi.org/10.1080/00018732.2016.1198134} {\bibfield  {journal}
  {\bibinfo  {journal} {Adv. Phys.}\ }\textbf {\bibinfo {volume} {65}},\
  \bibinfo {pages} {239} (\bibinfo {year} {2016})}\BibitemShut {NoStop}%
\bibitem [{\citenamefont {Deutsch}(1991)}]{deutsch1991}%
  \BibitemOpen
  \bibfield  {author} {\bibinfo {author} {\bibfnamefont {J.~M.}\ \bibnamefont
  {Deutsch}},\ }\bibfield  {title} {\bibinfo {title} {{Quantum statistical
  mechanics in a closed system}},\ }\href
  {https://doi.org/10.1103/PhysRevA.43.2046} {\bibfield  {journal} {\bibinfo
  {journal} {Phys. Rev. A}\ }\textbf {\bibinfo {volume} {43}},\ \bibinfo
  {pages} {2046} (\bibinfo {year} {1991})}\BibitemShut {NoStop}%
\bibitem [{\citenamefont {Srednicki}(1994)}]{srednicki1994}%
  \BibitemOpen
  \bibfield  {author} {\bibinfo {author} {\bibfnamefont {M.}~\bibnamefont
  {Srednicki}},\ }\bibfield  {title} {\bibinfo {title} {{Chaos and quantum
  thermalization}},\ }\href {https://doi.org/10.1103/PhysRevE.50.888}
  {\bibfield  {journal} {\bibinfo  {journal} {Phys. Rev. E}\ }\textbf {\bibinfo
  {volume} {50}},\ \bibinfo {pages} {888} (\bibinfo {year} {1994})}\BibitemShut
  {NoStop}%
\bibitem [{\citenamefont {Rigol}\ \emph {et~al.}(2008)\citenamefont {Rigol},
  \citenamefont {Dunjko},\ and\ \citenamefont {Olshanii}}]{rigol2008}%
  \BibitemOpen
  \bibfield  {author} {\bibinfo {author} {\bibfnamefont {M.}~\bibnamefont
  {Rigol}}, \bibinfo {author} {\bibfnamefont {V.}~\bibnamefont {Dunjko}},\ and\
  \bibinfo {author} {\bibfnamefont {M.}~\bibnamefont {Olshanii}},\ }\bibfield
  {title} {\bibinfo {title} {{Thermalization and its mechanism for generic
  isolated quantum systems}},\ }\href {https://doi.org/10.1038/nature06838}
  {\bibfield  {journal} {\bibinfo  {journal} {Nature}\ }\textbf {\bibinfo
  {volume} {452}},\ \bibinfo {pages} {854} (\bibinfo {year}
  {2008})}\BibitemShut {NoStop}%
\bibitem [{\citenamefont {Deutsch}(2018)}]{deutsch2018}%
  \BibitemOpen
  \bibfield  {author} {\bibinfo {author} {\bibfnamefont {J.~M.}\ \bibnamefont
  {Deutsch}},\ }\bibfield  {title} {\bibinfo {title} {{Eigenstate
  thermalization hypothesis}},\ }\href
  {https://dx.doi.org/10.1088/1361-6633/aac9f1} {\bibfield  {journal} {\bibinfo
   {journal} {Rep. Prog. Phys.}\ }\textbf {\bibinfo {volume} {81}},\ \bibinfo
  {pages} {082001} (\bibinfo {year} {2018})}\BibitemShut {NoStop}%
\bibitem [{\citenamefont {Alba}\ and\ \citenamefont
  {Fagotti}(2017)}]{alba2017}%
  \BibitemOpen
  \bibfield  {author} {\bibinfo {author} {\bibfnamefont {V.}~\bibnamefont
  {Alba}}\ and\ \bibinfo {author} {\bibfnamefont {M.}~\bibnamefont {Fagotti}},\
  }\bibfield  {title} {\bibinfo {title} {Prethermalization at low temperature:
  The scent of long-range order},\ }\href
  {https://doi.org/10.1103/PhysRevLett.119.010601} {\bibfield  {journal}
  {\bibinfo  {journal} {Phys. Rev. Lett.}\ }\textbf {\bibinfo {volume} {119}},\
  \bibinfo {pages} {010601} (\bibinfo {year} {2017})}\BibitemShut {NoStop}%
\bibitem [{\citenamefont {Fagotti}\ \emph {et~al.}(2022)\citenamefont
  {Fagotti}, \citenamefont {Mari{\'c}},\ and\ \citenamefont
  {Zadnik}}]{maric2022}%
  \BibitemOpen
  \bibfield  {author} {\bibinfo {author} {\bibfnamefont {M.}~\bibnamefont
  {Fagotti}}, \bibinfo {author} {\bibfnamefont {V.}~\bibnamefont {Mari{\'c}}},\
  and\ \bibinfo {author} {\bibfnamefont {L.}~\bibnamefont {Zadnik}},\
  }\bibfield  {title} {\bibinfo {title} {Nonequilibrium symmetry-protected
  topological order: emergence of semilocal gibbs ensembles},\ }\href
  {https://doi.org/10.48550/arXiv.2205.02221} {\bibfield  {journal} {\bibinfo
  {journal} {arXiv:2205.02221}\ } (\bibinfo {year} {2022})}\BibitemShut
  {NoStop}%
\bibitem [{\citenamefont {Ares}\ \emph {et~al.}(2023)\citenamefont {Ares},
  \citenamefont {Murciano},\ and\ \citenamefont {Calabrese}}]{ares2023}%
  \BibitemOpen
  \bibfield  {author} {\bibinfo {author} {\bibfnamefont {F.}~\bibnamefont
  {Ares}}, \bibinfo {author} {\bibfnamefont {S.}~\bibnamefont {Murciano}},\
  and\ \bibinfo {author} {\bibfnamefont {P.}~\bibnamefont {Calabrese}},\
  }\bibfield  {title} {\bibinfo {title} {{Entanglement asymmetry as a probe of
  symmetry breaking}},\ }\href {https://doi.org/10.1038/s41467-023-37747-8}
  {\bibfield  {journal} {\bibinfo  {journal} {Nature Communications}\ }\textbf
  {\bibinfo {volume} {14}},\ \bibinfo {pages} {2036} (\bibinfo {year}
  {2023})}\BibitemShut {NoStop}%
\bibitem [{\citenamefont {Fagotti}(2014)}]{fagotti2014}%
  \BibitemOpen
  \bibfield  {author} {\bibinfo {author} {\bibfnamefont {M.}~\bibnamefont
  {Fagotti}},\ }\bibfield  {title} {\bibinfo {title} {On conservation laws,
  relaxation and pre-relaxation after a quantum quench},\ }\href
  {https://doi.org/10.1088/1742-5468/2014/03/P03016} {\bibfield  {journal}
  {\bibinfo  {journal} {J. Stat. Phys.}\ }\textbf {\bibinfo {volume} {2014}},\
  \bibinfo {pages} {P03016} (\bibinfo {year} {2014})}\BibitemShut {NoStop}%
\bibitem [{\citenamefont {Bertini}\ and\ \citenamefont
  {Fagotti}(2015)}]{bertini2015}%
  \BibitemOpen
  \bibfield  {author} {\bibinfo {author} {\bibfnamefont {B.}~\bibnamefont
  {Bertini}}\ and\ \bibinfo {author} {\bibfnamefont {M.}~\bibnamefont
  {Fagotti}},\ }\bibfield  {title} {\bibinfo {title} {Pre-relaxation in weakly
  interacting models},\ }\href
  {https://dx.doi.org/10.1088/1742-5468/2015/07/P07012} {\bibfield  {journal}
  {\bibinfo  {journal} {J. Stat. Mech.}\ }\textbf {\bibinfo {volume} {2015}},\
  \bibinfo {pages} {P07012} (\bibinfo {year} {2015})}\BibitemShut {NoStop}%
\bibitem [{\citenamefont {Kemp}\ \emph {et~al.}(2017)\citenamefont {Kemp},
  \citenamefont {Yao}, \citenamefont {Laumann},\ and\ \citenamefont
  {Fendley}}]{kemp2017}%
  \BibitemOpen
  \bibfield  {author} {\bibinfo {author} {\bibfnamefont {J.}~\bibnamefont
  {Kemp}}, \bibinfo {author} {\bibfnamefont {N.~Y.}\ \bibnamefont {Yao}},
  \bibinfo {author} {\bibfnamefont {C.~R.}\ \bibnamefont {Laumann}},\ and\
  \bibinfo {author} {\bibfnamefont {P.}~\bibnamefont {Fendley}},\ }\bibfield
  {title} {\bibinfo {title} {Long coherence times for edge spins},\ }\href
  {https://doi.org/10.1088/1742-5468/aa73f0} {\bibfield  {journal} {\bibinfo
  {journal} {J. Stat. Mech.}\ }\textbf {\bibinfo {volume} {2017}},\ \bibinfo
  {pages} {063105} (\bibinfo {year} {2017})}\BibitemShut {NoStop}%
\bibitem [{\citenamefont {Abanin}\ \emph {et~al.}(2017)\citenamefont {Abanin},
  \citenamefont {De~Roeck}, \citenamefont {Ho},\ and\ \citenamefont
  {Huveneers}}]{abanin2017}%
  \BibitemOpen
  \bibfield  {author} {\bibinfo {author} {\bibfnamefont {D.}~\bibnamefont
  {Abanin}}, \bibinfo {author} {\bibfnamefont {W.}~\bibnamefont {De~Roeck}},
  \bibinfo {author} {\bibfnamefont {W.~W.}\ \bibnamefont {Ho}},\ and\ \bibinfo
  {author} {\bibfnamefont {F.}~\bibnamefont {Huveneers}},\ }\bibfield  {title}
  {\bibinfo {title} {A rigorous theory of many-body prethermalization for
  periodically driven and closed quantum systems},\ }\href
  {https://doi.org/10.1007/s00220-017-2930-x} {\bibfield  {journal} {\bibinfo
  {journal} {Comm. Math. Phys.}\ }\textbf {\bibinfo {volume} {354}},\ \bibinfo
  {pages} {809} (\bibinfo {year} {2017})}\BibitemShut {NoStop}%
\bibitem [{\citenamefont {Mori}\ \emph {et~al.}(2018)\citenamefont {Mori},
  \citenamefont {Ikeda}, \citenamefont {Kaminishi},\ and\ \citenamefont
  {Ueda}}]{mori2018}%
  \BibitemOpen
  \bibfield  {author} {\bibinfo {author} {\bibfnamefont {T.}~\bibnamefont
  {Mori}}, \bibinfo {author} {\bibfnamefont {T.~N.}\ \bibnamefont {Ikeda}},
  \bibinfo {author} {\bibfnamefont {E.}~\bibnamefont {Kaminishi}},\ and\
  \bibinfo {author} {\bibfnamefont {M.}~\bibnamefont {Ueda}},\ }\bibfield
  {title} {\bibinfo {title} {Thermalization and prethermalization in isolated
  quantum systems: a theoretical overview},\ }\href
  {https://dx.doi.org/10.1088/1361-6455/aabcdf} {\bibfield  {journal} {\bibinfo
   {journal} {J. Phys. B}\ }\textbf {\bibinfo {volume} {51}},\ \bibinfo {pages}
  {112001} (\bibinfo {year} {2018})}\BibitemShut {NoStop}%
\bibitem [{\citenamefont {van Horssen}\ \emph {et~al.}(2015)\citenamefont {van
  Horssen}, \citenamefont {Levi},\ and\ \citenamefont
  {Garrahan}}]{horssen2015}%
  \BibitemOpen
  \bibfield  {author} {\bibinfo {author} {\bibfnamefont {M.}~\bibnamefont {van
  Horssen}}, \bibinfo {author} {\bibfnamefont {E.}~\bibnamefont {Levi}},\ and\
  \bibinfo {author} {\bibfnamefont {J.~P.}\ \bibnamefont {Garrahan}},\
  }\bibfield  {title} {\bibinfo {title} {Dynamics of many-body localization in
  a translation-invariant quantum glass model},\ }\href
  {https://doi.org/10.1103/PhysRevB.92.100305} {\bibfield  {journal} {\bibinfo
  {journal} {Phys. Rev. B}\ }\textbf {\bibinfo {volume} {92}},\ \bibinfo
  {pages} {100305} (\bibinfo {year} {2015})}\BibitemShut {NoStop}%
\bibitem [{\citenamefont {Lan}\ \emph {et~al.}(2018)\citenamefont {Lan},
  \citenamefont {van Horssen}, \citenamefont {Powell},\ and\ \citenamefont
  {Garrahan}}]{lan2018}%
  \BibitemOpen
  \bibfield  {author} {\bibinfo {author} {\bibfnamefont {Z.}~\bibnamefont
  {Lan}}, \bibinfo {author} {\bibfnamefont {M.}~\bibnamefont {van Horssen}},
  \bibinfo {author} {\bibfnamefont {S.}~\bibnamefont {Powell}},\ and\ \bibinfo
  {author} {\bibfnamefont {J.~P.}\ \bibnamefont {Garrahan}},\ }\bibfield
  {title} {\bibinfo {title} {Quantum slow relaxation and metastability due to
  dynamical constraints},\ }\href
  {https://doi.org/10.1103/PhysRevLett.121.040603} {\bibfield  {journal}
  {\bibinfo  {journal} {Phys. Rev. Lett.}\ }\textbf {\bibinfo {volume} {121}},\
  \bibinfo {pages} {040603} (\bibinfo {year} {2018})}\BibitemShut {NoStop}%
\bibitem [{\citenamefont {Morningstar}\ \emph {et~al.}(2020)\citenamefont
  {Morningstar}, \citenamefont {Khemani},\ and\ \citenamefont
  {Huse}}]{morningstar2020}%
  \BibitemOpen
  \bibfield  {author} {\bibinfo {author} {\bibfnamefont {A.}~\bibnamefont
  {Morningstar}}, \bibinfo {author} {\bibfnamefont {V.}~\bibnamefont
  {Khemani}},\ and\ \bibinfo {author} {\bibfnamefont {D.~A.}\ \bibnamefont
  {Huse}},\ }\bibfield  {title} {\bibinfo {title} {Kinetically constrained
  freezing transition in a dipole-conserving system},\ }\href
  {https://doi.org/10.1103/PhysRevB.101.214205} {\bibfield  {journal} {\bibinfo
   {journal} {Phys. Rev. B}\ }\textbf {\bibinfo {volume} {101}},\ \bibinfo
  {pages} {214205} (\bibinfo {year} {2020})}\BibitemShut {NoStop}%
\bibitem [{\citenamefont {Pancotti}\ \emph {et~al.}(2020)\citenamefont
  {Pancotti}, \citenamefont {Giudice}, \citenamefont {Cirac}, \citenamefont
  {Garrahan},\ and\ \citenamefont {Ba\~nuls}}]{pancotti2020}%
  \BibitemOpen
  \bibfield  {author} {\bibinfo {author} {\bibfnamefont {N.}~\bibnamefont
  {Pancotti}}, \bibinfo {author} {\bibfnamefont {G.}~\bibnamefont {Giudice}},
  \bibinfo {author} {\bibfnamefont {J.~I.}\ \bibnamefont {Cirac}}, \bibinfo
  {author} {\bibfnamefont {J.~P.}\ \bibnamefont {Garrahan}},\ and\ \bibinfo
  {author} {\bibfnamefont {M.~C.}\ \bibnamefont {Ba\~nuls}},\ }\bibfield
  {title} {\bibinfo {title} {Quantum east model: Localization, nonthermal
  eigenstates, and slow dynamics},\ }\href
  {https://doi.org/10.1103/PhysRevX.10.021051} {\bibfield  {journal} {\bibinfo
  {journal} {Phys. Rev. X}\ }\textbf {\bibinfo {volume} {10}},\ \bibinfo
  {pages} {021051} (\bibinfo {year} {2020})}\BibitemShut {NoStop}%
\bibitem [{\citenamefont {Scherg}\ \emph {et~al.}(2021)\citenamefont {Scherg},
  \citenamefont {Kohlert}, \citenamefont {Sala}, \citenamefont {Pollmann},
  \citenamefont {Hebbe~Madhusudhana}, \citenamefont {Bloch},\ and\
  \citenamefont {Aidelsburger}}]{scherg2021}%
  \BibitemOpen
  \bibfield  {author} {\bibinfo {author} {\bibfnamefont {S.}~\bibnamefont
  {Scherg}}, \bibinfo {author} {\bibfnamefont {T.}~\bibnamefont {Kohlert}},
  \bibinfo {author} {\bibfnamefont {P.}~\bibnamefont {Sala}}, \bibinfo {author}
  {\bibfnamefont {F.}~\bibnamefont {Pollmann}}, \bibinfo {author}
  {\bibfnamefont {B.}~\bibnamefont {Hebbe~Madhusudhana}}, \bibinfo {author}
  {\bibfnamefont {I.}~\bibnamefont {Bloch}},\ and\ \bibinfo {author}
  {\bibfnamefont {M.}~\bibnamefont {Aidelsburger}},\ }\bibfield  {title}
  {\bibinfo {title} {Observing non-ergodicity due to kinetic constraints in
  tilted fermi-hubbard chains},\ }\href
  {https://doi.org/10.1038/s41467-021-24726-0} {\bibfield  {journal} {\bibinfo
  {journal} {Nature Comm.}\ }\textbf {\bibinfo {volume} {12}},\ \bibinfo
  {pages} {4490} (\bibinfo {year} {2021})}\BibitemShut {NoStop}%
\bibitem [{\citenamefont {Brighi}\ \emph {et~al.}(2022)\citenamefont {Brighi},
  \citenamefont {Ljubotina},\ and\ \citenamefont {Serbyn}}]{brighi2022}%
  \BibitemOpen
  \bibfield  {author} {\bibinfo {author} {\bibfnamefont {P.}~\bibnamefont
  {Brighi}}, \bibinfo {author} {\bibfnamefont {M.}~\bibnamefont {Ljubotina}},\
  and\ \bibinfo {author} {\bibfnamefont {M.}~\bibnamefont {Serbyn}},\
  }\bibfield  {title} {\bibinfo {title} {Hilbert space fragmentation and slow
  dynamics in particle-conserving quantum east models},\ }\href
  {https://doi.org/10.48550/arXiv.2210.15607} {\bibfield  {journal} {\bibinfo
  {journal} {arXiv:2210.15607}\ } (\bibinfo {year} {2022})}\BibitemShut
  {NoStop}%
\bibitem [{\citenamefont {Valencia-Tortora}\ \emph {et~al.}(2022)\citenamefont
  {Valencia-Tortora}, \citenamefont {Pancotti},\ and\ \citenamefont
  {Marino}}]{tortora2022}%
  \BibitemOpen
  \bibfield  {author} {\bibinfo {author} {\bibfnamefont {R.~J.}\ \bibnamefont
  {Valencia-Tortora}}, \bibinfo {author} {\bibfnamefont {N.}~\bibnamefont
  {Pancotti}},\ and\ \bibinfo {author} {\bibfnamefont {J.}~\bibnamefont
  {Marino}},\ }\bibfield  {title} {\bibinfo {title} {Kinetically constrained
  quantum dynamics in superconducting circuits},\ }\href
  {https://link.aps.org/doi/10.1103/PRXQuantum.3.020346} {\bibfield  {journal}
  {\bibinfo  {journal} {PRX Quantum}\ }\textbf {\bibinfo {volume} {3}},\
  \bibinfo {pages} {020346} (\bibinfo {year} {2022})}\BibitemShut {NoStop}%
\bibitem [{\citenamefont {Deger}\ \emph
  {et~al.}(2022{\natexlab{a}})\citenamefont {Deger}, \citenamefont {Roy},\ and\
  \citenamefont {Lazarides}}]{deger2022arresting}%
  \BibitemOpen
  \bibfield  {author} {\bibinfo {author} {\bibfnamefont {A.}~\bibnamefont
  {Deger}}, \bibinfo {author} {\bibfnamefont {S.}~\bibnamefont {Roy}},\ and\
  \bibinfo {author} {\bibfnamefont {A.}~\bibnamefont {Lazarides}},\ }\bibfield
  {title} {\bibinfo {title} {Arresting classical many-body chaos by kinetic
  constraints},\ }\href {https://doi.org/10.1103/PhysRevLett.129.160601}
  {\bibfield  {journal} {\bibinfo  {journal} {Phys. Rev. Lett.}\ }\textbf
  {\bibinfo {volume} {129}},\ \bibinfo {pages} {160601} (\bibinfo {year}
  {2022}{\natexlab{a}})}\BibitemShut {NoStop}%
\bibitem [{\citenamefont {Deger}\ \emph
  {et~al.}(2022{\natexlab{b}})\citenamefont {Deger}, \citenamefont
  {Lazarides},\ and\ \citenamefont {Roy}}]{deger2022constrained}%
  \BibitemOpen
  \bibfield  {author} {\bibinfo {author} {\bibfnamefont {A.}~\bibnamefont
  {Deger}}, \bibinfo {author} {\bibfnamefont {A.}~\bibnamefont {Lazarides}},\
  and\ \bibinfo {author} {\bibfnamefont {S.}~\bibnamefont {Roy}},\ }\bibfield
  {title} {\bibinfo {title} {Constrained dynamics and directed percolation},\
  }\href {https://doi.org/10.1103/PhysRevLett.129.190601} {\bibfield  {journal}
  {\bibinfo  {journal} {Phys. Rev. Lett.}\ }\textbf {\bibinfo {volume} {129}},\
  \bibinfo {pages} {190601} (\bibinfo {year} {2022}{\natexlab{b}})}\BibitemShut
  {NoStop}%
\bibitem [{\citenamefont {Palmer}\ \emph {et~al.}(1984)\citenamefont {Palmer},
  \citenamefont {Stein}, \citenamefont {Abrahams},\ and\ \citenamefont
  {Anderson}}]{palmer1984}%
  \BibitemOpen
  \bibfield  {author} {\bibinfo {author} {\bibfnamefont {R.~G.}\ \bibnamefont
  {Palmer}}, \bibinfo {author} {\bibfnamefont {D.~L.}\ \bibnamefont {Stein}},
  \bibinfo {author} {\bibfnamefont {E.}~\bibnamefont {Abrahams}},\ and\
  \bibinfo {author} {\bibfnamefont {P.~W.}\ \bibnamefont {Anderson}},\
  }\bibfield  {title} {\bibinfo {title} {Models of hierarchically constrained
  dynamics for glassy relaxation},\ }\href
  {https://doi.org/10.1103/PhysRevLett.53.958} {\bibfield  {journal} {\bibinfo
  {journal} {Phys. Rev. Lett.}\ }\textbf {\bibinfo {volume} {53}},\ \bibinfo
  {pages} {958} (\bibinfo {year} {1984})}\BibitemShut {NoStop}%
\bibitem [{\citenamefont {Fredrickson}\ and\ \citenamefont
  {Andersen}(1984)}]{fredrickson1984}%
  \BibitemOpen
  \bibfield  {author} {\bibinfo {author} {\bibfnamefont {G.~H.}\ \bibnamefont
  {Fredrickson}}\ and\ \bibinfo {author} {\bibfnamefont {H.~C.}\ \bibnamefont
  {Andersen}},\ }\bibfield  {title} {\bibinfo {title} {Kinetic ising model of
  the glass transition},\ }\href {https://doi.org/10.1103/PhysRevLett.53.1244}
  {\bibfield  {journal} {\bibinfo  {journal} {Phys. Rev. Lett.}\ }\textbf
  {\bibinfo {volume} {53}},\ \bibinfo {pages} {1244} (\bibinfo {year}
  {1984})}\BibitemShut {NoStop}%
\bibitem [{\citenamefont {Ritort}\ and\ \citenamefont
  {Sollich}(2003)}]{ritort2003}%
  \BibitemOpen
  \bibfield  {author} {\bibinfo {author} {\bibfnamefont {F.}~\bibnamefont
  {Ritort}}\ and\ \bibinfo {author} {\bibfnamefont {P.}~\bibnamefont
  {Sollich}},\ }\bibfield  {title} {\bibinfo {title} {Glassy dynamics of
  kinetically constrained models},\ }\href
  {https://doi.org/10.1080/0001873031000093582} {\bibfield  {journal} {\bibinfo
   {journal} {Adv. Phys.}\ }\textbf {\bibinfo {volume} {52}},\ \bibinfo {pages}
  {219} (\bibinfo {year} {2003})}\BibitemShut {NoStop}%
\bibitem [{\citenamefont {Garrahan}(2018)}]{garrahan2018aspects}%
  \BibitemOpen
  \bibfield  {author} {\bibinfo {author} {\bibfnamefont {J.~P.}\ \bibnamefont
  {Garrahan}},\ }\bibfield  {title} {\bibinfo {title} {Aspects of
  non-equilibrium in classical and quantum systems: Slow relaxation and
  glasses, dynamical large deviations, quantum non-ergodicity, and open quantum
  dynamics},\ }\href
  {https://doi.org/https://doi.org/10.1016/j.physa.2017.12.149} {\bibfield
  {journal} {\bibinfo  {journal} {Physica A}\ }\textbf {\bibinfo {volume}
  {504}},\ \bibinfo {pages} {130} (\bibinfo {year} {2018})}\BibitemShut
  {NoStop}%
\bibitem [{\citenamefont {Chandler}\ and\ \citenamefont
  {Garrahan}(2010)}]{chandler2010dynamics}%
  \BibitemOpen
  \bibfield  {author} {\bibinfo {author} {\bibfnamefont {D.}~\bibnamefont
  {Chandler}}\ and\ \bibinfo {author} {\bibfnamefont {J.~P.}\ \bibnamefont
  {Garrahan}},\ }\bibfield  {title} {\bibinfo {title} {Dynamics on the way to
  forming glass: Bubbles in space-time},\ }\href
  {https://doi.org/10.1146/annurev.physchem.040808.090405} {\bibfield
  {journal} {\bibinfo  {journal} {Annu. Rev. Phys. Chem.}\ }\textbf {\bibinfo
  {volume} {61}},\ \bibinfo {pages} {191} (\bibinfo {year} {2010})}\BibitemShut
  {NoStop}%
\bibitem [{\citenamefont {Berthier}\ and\ \citenamefont
  {Biroli}(2011)}]{berthier2011theoretical}%
  \BibitemOpen
  \bibfield  {author} {\bibinfo {author} {\bibfnamefont {L.}~\bibnamefont
  {Berthier}}\ and\ \bibinfo {author} {\bibfnamefont {G.}~\bibnamefont
  {Biroli}},\ }\bibfield  {title} {\bibinfo {title} {{Theoretical perspective
  on the glass transition and amorphous materials}},\ }\href
  {https://link.aps.org/doi/10.1103/RevModPhys.83.587} {\bibfield  {journal}
  {\bibinfo  {journal} {Rev. Mod. Phys.}\ }\textbf {\bibinfo {volume} {83}},\
  \bibinfo {pages} {587} (\bibinfo {year} {2011})}\BibitemShut {NoStop}%
\bibitem [{\citenamefont {Biroli}\ and\ \citenamefont
  {Garrahan}(2013)}]{biroli2013perspective}%
  \BibitemOpen
  \bibfield  {author} {\bibinfo {author} {\bibfnamefont {G.}~\bibnamefont
  {Biroli}}\ and\ \bibinfo {author} {\bibfnamefont {J.~P.}\ \bibnamefont
  {Garrahan}},\ }\bibfield  {title} {\bibinfo {title} {{Perspective: The glass
  transition}},\ }\href {https://doi.org/10.1063/1.4795539} {\bibfield
  {journal} {\bibinfo  {journal} {J. Chem. Phys.}\ }\textbf {\bibinfo {volume}
  {138}},\ \bibinfo {eid} {12A301} (\bibinfo {year} {2013})}\BibitemShut
  {NoStop}%
\bibitem [{\citenamefont {Lesanovsky}(2011)}]{lesanovsky2011}%
  \BibitemOpen
  \bibfield  {author} {\bibinfo {author} {\bibfnamefont {I.}~\bibnamefont
  {Lesanovsky}},\ }\bibfield  {title} {\bibinfo {title} {{Many-Body spin
  interactions and the ground state of a dense Rydberg lattice gas}},\ }\href
  {https://doi.org/10.1103/PhysRevLett.106.025301} {\bibfield  {journal}
  {\bibinfo  {journal} {Phys. Rev. Lett.}\ }\textbf {\bibinfo {volume} {106}},\
  \bibinfo {pages} {025301} (\bibinfo {year} {2011})}\BibitemShut {NoStop}%
\bibitem [{\citenamefont {Browaeys}\ and\ \citenamefont
  {Lahaye}(2020)}]{browaeys2020many-body}%
  \BibitemOpen
  \bibfield  {author} {\bibinfo {author} {\bibfnamefont {A.}~\bibnamefont
  {Browaeys}}\ and\ \bibinfo {author} {\bibfnamefont {T.}~\bibnamefont
  {Lahaye}},\ }\bibfield  {title} {\bibinfo {title} {{Many-body physics with
  individually controlled Rydberg atoms}},\ }\href
  {https://doi.org/10.1038/s41567-019-0733-z} {\bibfield  {journal} {\bibinfo
  {journal} {Nature Phys.}\ }\textbf {\bibinfo {volume} {16}},\ \bibinfo
  {pages} {132} (\bibinfo {year} {2020})}\BibitemShut {NoStop}%
\bibitem [{\citenamefont {Yang}\ \emph {et~al.}(2020)\citenamefont {Yang},
  \citenamefont {Liu}, \citenamefont {Gorshkov},\ and\ \citenamefont
  {Iadecola}}]{yang2020}%
  \BibitemOpen
  \bibfield  {author} {\bibinfo {author} {\bibfnamefont {Z.-C.}\ \bibnamefont
  {Yang}}, \bibinfo {author} {\bibfnamefont {F.}~\bibnamefont {Liu}}, \bibinfo
  {author} {\bibfnamefont {A.~V.}\ \bibnamefont {Gorshkov}},\ and\ \bibinfo
  {author} {\bibfnamefont {T.}~\bibnamefont {Iadecola}},\ }\bibfield  {title}
  {\bibinfo {title} {Hilbert-space fragmentation from strict confinement},\
  }\href {https://doi.org/10.1103/PhysRevLett.124.207602} {\bibfield  {journal}
  {\bibinfo  {journal} {Phys. Rev. Lett.}\ }\textbf {\bibinfo {volume} {124}},\
  \bibinfo {pages} {207602} (\bibinfo {year} {2020})}\BibitemShut {NoStop}%
\bibitem [{\citenamefont {Langlett}\ and\ \citenamefont
  {Xu}(2021)}]{langlett2021}%
  \BibitemOpen
  \bibfield  {author} {\bibinfo {author} {\bibfnamefont {C.~M.}\ \bibnamefont
  {Langlett}}\ and\ \bibinfo {author} {\bibfnamefont {S.}~\bibnamefont {Xu}},\
  }\bibfield  {title} {\bibinfo {title} {{Hilbert space fragmentation and exact
  scars of generalized Fredkin spin chains}},\ }\href
  {https://doi.org/10.1103/PhysRevB.103.L220304} {\bibfield  {journal}
  {\bibinfo  {journal} {Phys. Rev. B}\ }\textbf {\bibinfo {volume} {103}},\
  \bibinfo {pages} {L220304} (\bibinfo {year} {2021})}\BibitemShut {NoStop}%
\bibitem [{\citenamefont {Zadnik}\ and\ \citenamefont
  {Fagotti}(2021)}]{zadnik2021}%
  \BibitemOpen
  \bibfield  {author} {\bibinfo {author} {\bibfnamefont {L.}~\bibnamefont
  {Zadnik}}\ and\ \bibinfo {author} {\bibfnamefont {M.}~\bibnamefont
  {Fagotti}},\ }\bibfield  {title} {\bibinfo {title} {{The folded spin-1/2 XXZ
  model: I. Diagonalisation, jamming, and ground state properties}},\ }\href
  {https://doi.org/10.21468/SciPostPhysCore.4.2.010} {\bibfield  {journal}
  {\bibinfo  {journal} {SciPost Phys. Core}\ }\textbf {\bibinfo {volume} {4}},\
  \bibinfo {pages} {010} (\bibinfo {year} {2021})}\BibitemShut {NoStop}%
\bibitem [{\citenamefont {Pozsgay}\ \emph {et~al.}(2021)\citenamefont
  {Pozsgay}, \citenamefont {Gombor}, \citenamefont {Hutsalyuk}, \citenamefont
  {Jiang}, \citenamefont {Pristy\'ak},\ and\ \citenamefont
  {Vernier}}]{pozsgay2021}%
  \BibitemOpen
  \bibfield  {author} {\bibinfo {author} {\bibfnamefont {B.}~\bibnamefont
  {Pozsgay}}, \bibinfo {author} {\bibfnamefont {T.}~\bibnamefont {Gombor}},
  \bibinfo {author} {\bibfnamefont {A.}~\bibnamefont {Hutsalyuk}}, \bibinfo
  {author} {\bibfnamefont {Y.}~\bibnamefont {Jiang}}, \bibinfo {author}
  {\bibfnamefont {L.}~\bibnamefont {Pristy\'ak}},\ and\ \bibinfo {author}
  {\bibfnamefont {E.}~\bibnamefont {Vernier}},\ }\bibfield  {title} {\bibinfo
  {title} {{Integrable spin chain with Hilbert space fragmentation and solvable
  real-time dynamics}},\ }\href {https://doi.org/10.1103/PhysRevE.104.044106}
  {\bibfield  {journal} {\bibinfo  {journal} {Phys. Rev. E}\ }\textbf {\bibinfo
  {volume} {104}},\ \bibinfo {pages} {044106} (\bibinfo {year}
  {2021})}\BibitemShut {NoStop}%
\bibitem [{\citenamefont {Tamura}\ and\ \citenamefont
  {Katsura}(2022)}]{tamura2022}%
  \BibitemOpen
  \bibfield  {author} {\bibinfo {author} {\bibfnamefont {K.}~\bibnamefont
  {Tamura}}\ and\ \bibinfo {author} {\bibfnamefont {H.}~\bibnamefont
  {Katsura}},\ }\bibfield  {title} {\bibinfo {title} {Quantum many-body scars
  of spinless fermions with density-assisted hopping in higher dimensions},\
  }\href {https://doi.org/10.1103/PhysRevB.106.144306} {\bibfield  {journal}
  {\bibinfo  {journal} {Phys. Rev. B}\ }\textbf {\bibinfo {volume} {106}},\
  \bibinfo {pages} {144306} (\bibinfo {year} {2022})}\BibitemShut {NoStop}%
\bibitem [{\citenamefont {Bidzhiev}\ \emph {et~al.}(2022)\citenamefont
  {Bidzhiev}, \citenamefont {Fagotti},\ and\ \citenamefont
  {Zadnik}}]{bidzhiev2022}%
  \BibitemOpen
  \bibfield  {author} {\bibinfo {author} {\bibfnamefont {K.}~\bibnamefont
  {Bidzhiev}}, \bibinfo {author} {\bibfnamefont {M.}~\bibnamefont {Fagotti}},\
  and\ \bibinfo {author} {\bibfnamefont {L.}~\bibnamefont {Zadnik}},\
  }\bibfield  {title} {\bibinfo {title} {Macroscopic effects of localized
  measurements in jammed states of quantum spin chains},\ }\href
  {https://doi.org/10.1103/PhysRevLett.128.130603} {\bibfield  {journal}
  {\bibinfo  {journal} {Phys. Rev. Lett.}\ }\textbf {\bibinfo {volume} {128}},\
  \bibinfo {pages} {130603} (\bibinfo {year} {2022})}\BibitemShut {NoStop}%
\bibitem [{\citenamefont {Bernien}\ \emph {et~al.}(2017)\citenamefont
  {Bernien}, \citenamefont {Schwartz}, \citenamefont {Keesling}, \citenamefont
  {Levine}, \citenamefont {Omran}, \citenamefont {Pichler}, \citenamefont
  {Choi}, \citenamefont {Zibrov}, \citenamefont {Endres}, \citenamefont
  {Greiner}, \citenamefont {Vuleti{\'{c}}},\ and\ \citenamefont
  {Lukin}}]{bernien2017}%
  \BibitemOpen
  \bibfield  {author} {\bibinfo {author} {\bibfnamefont {H.}~\bibnamefont
  {Bernien}}, \bibinfo {author} {\bibfnamefont {S.}~\bibnamefont {Schwartz}},
  \bibinfo {author} {\bibfnamefont {A.}~\bibnamefont {Keesling}}, \bibinfo
  {author} {\bibfnamefont {H.}~\bibnamefont {Levine}}, \bibinfo {author}
  {\bibfnamefont {A.}~\bibnamefont {Omran}}, \bibinfo {author} {\bibfnamefont
  {H.}~\bibnamefont {Pichler}}, \bibinfo {author} {\bibfnamefont
  {S.}~\bibnamefont {Choi}}, \bibinfo {author} {\bibfnamefont {A.~S.}\
  \bibnamefont {Zibrov}}, \bibinfo {author} {\bibfnamefont {M.}~\bibnamefont
  {Endres}}, \bibinfo {author} {\bibfnamefont {M.}~\bibnamefont {Greiner}},
  \bibinfo {author} {\bibfnamefont {V.}~\bibnamefont {Vuleti{\'{c}}}},\ and\
  \bibinfo {author} {\bibfnamefont {M.~D.}\ \bibnamefont {Lukin}},\ }\bibfield
  {title} {\bibinfo {title} {{Probing many-body dynamics on a 51-atom quantum
  simulator}},\ }\href {https://doi.org/10.1038/nature24622} {\bibfield
  {journal} {\bibinfo  {journal} {Nature}\ }\textbf {\bibinfo {volume} {551}},\
  \bibinfo {pages} {579} (\bibinfo {year} {2017})}\BibitemShut {NoStop}%
\bibitem [{\citenamefont {Turner}\ \emph {et~al.}(2018)\citenamefont {Turner},
  \citenamefont {Michailidis}, \citenamefont {Abanin}, \citenamefont {Serbyn},\
  and\ \citenamefont {Papi{\'{c}}}}]{turner2018}%
  \BibitemOpen
  \bibfield  {author} {\bibinfo {author} {\bibfnamefont {C.~J.}\ \bibnamefont
  {Turner}}, \bibinfo {author} {\bibfnamefont {A.~A.}\ \bibnamefont
  {Michailidis}}, \bibinfo {author} {\bibfnamefont {D.~A.}\ \bibnamefont
  {Abanin}}, \bibinfo {author} {\bibfnamefont {M.}~\bibnamefont {Serbyn}},\
  and\ \bibinfo {author} {\bibfnamefont {Z.}~\bibnamefont {Papi{\'{c}}}},\
  }\bibfield  {title} {\bibinfo {title} {{Weak ergodicity breaking from quantum
  many-body scars}},\ }\href {https://doi.org/10.1038/s41567-018-0137-5}
  {\bibfield  {journal} {\bibinfo  {journal} {Nature Phys.}\ }\textbf {\bibinfo
  {volume} {14}},\ \bibinfo {pages} {745} (\bibinfo {year} {2018})}\BibitemShut
  {NoStop}%
\bibitem [{\citenamefont {Moudgalya}\ \emph {et~al.}(2022)\citenamefont
  {Moudgalya}, \citenamefont {Bernevig},\ and\ \citenamefont
  {Regnault}}]{moudgalya2022}%
  \BibitemOpen
  \bibfield  {author} {\bibinfo {author} {\bibfnamefont {S.}~\bibnamefont
  {Moudgalya}}, \bibinfo {author} {\bibfnamefont {B.~A.}\ \bibnamefont
  {Bernevig}},\ and\ \bibinfo {author} {\bibfnamefont {N.}~\bibnamefont
  {Regnault}},\ }\bibfield  {title} {\bibinfo {title} {{Quantum many-body scars
  and Hilbert space fragmentation: a review of exact results}},\ }\href
  {https://dx.doi.org/10.1088/1361-6633/ac73a0} {\bibfield  {journal} {\bibinfo
   {journal} {Rep. Prog. Phys.}\ }\textbf {\bibinfo {volume} {85}},\ \bibinfo
  {pages} {086501} (\bibinfo {year} {2022})}\BibitemShut {NoStop}%
\bibitem [{\citenamefont {Bluvstein}\ \emph {et~al.}(2021)\citenamefont
  {Bluvstein}, \citenamefont {Omran}, \citenamefont {Levine}, \citenamefont
  {Keesling}, \citenamefont {Semeghini}, \citenamefont {Ebadi}, \citenamefont
  {Wang}, \citenamefont {Michailidis}, \citenamefont {Maskara}, \citenamefont
  {Ho}, \citenamefont {Choi}, \citenamefont {Serbyn}, \citenamefont {Greiner},
  \citenamefont {Vuleti{\'c}},\ and\ \citenamefont {Lukin}}]{bluvstein2021}%
  \BibitemOpen
  \bibfield  {author} {\bibinfo {author} {\bibfnamefont {D.}~\bibnamefont
  {Bluvstein}}, \bibinfo {author} {\bibfnamefont {A.}~\bibnamefont {Omran}},
  \bibinfo {author} {\bibfnamefont {H.}~\bibnamefont {Levine}}, \bibinfo
  {author} {\bibfnamefont {A.}~\bibnamefont {Keesling}}, \bibinfo {author}
  {\bibfnamefont {G.}~\bibnamefont {Semeghini}}, \bibinfo {author}
  {\bibfnamefont {S.}~\bibnamefont {Ebadi}}, \bibinfo {author} {\bibfnamefont
  {T.~T.}\ \bibnamefont {Wang}}, \bibinfo {author} {\bibfnamefont {A.~A.}\
  \bibnamefont {Michailidis}}, \bibinfo {author} {\bibfnamefont
  {N.}~\bibnamefont {Maskara}}, \bibinfo {author} {\bibfnamefont {W.~W.}\
  \bibnamefont {Ho}}, \bibinfo {author} {\bibfnamefont {S.}~\bibnamefont
  {Choi}}, \bibinfo {author} {\bibfnamefont {M.}~\bibnamefont {Serbyn}},
  \bibinfo {author} {\bibfnamefont {M.}~\bibnamefont {Greiner}}, \bibinfo
  {author} {\bibfnamefont {V.}~\bibnamefont {Vuleti{\'c}}},\ and\ \bibinfo
  {author} {\bibfnamefont {M.~D.}\ \bibnamefont {Lukin}},\ }\bibfield  {title}
  {\bibinfo {title} {{Controlling quantum many-body dynamics in driven Rydberg
  atom arrays}},\ }\href {https://doi.org/10.1126/science.abg2530} {\bibfield
  {journal} {\bibinfo  {journal} {Science}\ }\textbf {\bibinfo {volume}
  {371}},\ \bibinfo {pages} {1355} (\bibinfo {year} {2021})}\BibitemShut
  {NoStop}%
\bibitem [{\citenamefont {Yang}(2022)}]{yang2022}%
  \BibitemOpen
  \bibfield  {author} {\bibinfo {author} {\bibfnamefont {Z.-C.}\ \bibnamefont
  {Yang}},\ }\bibfield  {title} {\bibinfo {title} {{Distinction Between
  Transport and R\'enyi Entropy Growth in Kinetically Constrained Models}},\
  }\href {https://doi.org/10.48550/arXiv.2208.07480} {\bibfield  {journal}
  {\bibinfo  {journal} {arXiv:2208.07480}\ } (\bibinfo {year}
  {2022})}\BibitemShut {NoStop}%
\bibitem [{\citenamefont {Ljubotina}\ \emph {et~al.}(2023)\citenamefont
  {Ljubotina}, \citenamefont {Desaules}, \citenamefont {Serbyn},\ and\
  \citenamefont {Papi\ifmmode~\acute{c}\else \'{c}\fi{}}}]{ljubotina2023}%
  \BibitemOpen
  \bibfield  {author} {\bibinfo {author} {\bibfnamefont {M.}~\bibnamefont
  {Ljubotina}}, \bibinfo {author} {\bibfnamefont {J.-Y.}\ \bibnamefont
  {Desaules}}, \bibinfo {author} {\bibfnamefont {M.}~\bibnamefont {Serbyn}},\
  and\ \bibinfo {author} {\bibfnamefont {Z.}~\bibnamefont
  {Papi\ifmmode~\acute{c}\else \'{c}\fi{}}},\ }\bibfield  {title} {\bibinfo
  {title} {Superdiffusive energy transport in kinetically constrained models},\
  }\href {https://doi.org/10.1103/PhysRevX.13.011033} {\bibfield  {journal}
  {\bibinfo  {journal} {Phys. Rev. X}\ }\textbf {\bibinfo {volume} {13}},\
  \bibinfo {pages} {011033} (\bibinfo {year} {2023})}\BibitemShut {NoStop}%
\bibitem [{\citenamefont {Nandkishore}\ and\ \citenamefont
  {Hermele}(2019)}]{nandkishore2019}%
  \BibitemOpen
  \bibfield  {author} {\bibinfo {author} {\bibfnamefont {R.~M.}\ \bibnamefont
  {Nandkishore}}\ and\ \bibinfo {author} {\bibfnamefont {M.}~\bibnamefont
  {Hermele}},\ }\bibfield  {title} {\bibinfo {title} {Fractons},\ }\href
  {https://doi.org/10.1146/annurev-conmatphys-031218-013604} {\bibfield
  {journal} {\bibinfo  {journal} {Ann. Rev. Condens. Matter}\ }\textbf
  {\bibinfo {volume} {10}},\ \bibinfo {pages} {295} (\bibinfo {year}
  {2019})}\BibitemShut {NoStop}%
\bibitem [{\citenamefont {Pretko}\ \emph {et~al.}(2020)\citenamefont {Pretko},
  \citenamefont {Chen},\ and\ \citenamefont {You}}]{pretko2020}%
  \BibitemOpen
  \bibfield  {author} {\bibinfo {author} {\bibfnamefont {M.}~\bibnamefont
  {Pretko}}, \bibinfo {author} {\bibfnamefont {X.}~\bibnamefont {Chen}},\ and\
  \bibinfo {author} {\bibfnamefont {Y.}~\bibnamefont {You}},\ }\bibfield
  {title} {\bibinfo {title} {Fracton phases of matter},\ }\href
  {https://doi.org/10.1142/S0217751X20300033} {\bibfield  {journal} {\bibinfo
  {journal} {Int. J. Mod. Phys. A}\ }\textbf {\bibinfo {volume} {35}},\
  \bibinfo {pages} {2030003} (\bibinfo {year} {2020})}\BibitemShut {NoStop}%
\bibitem [{\citenamefont {Izergin}\ \emph {et~al.}(1998)\citenamefont
  {Izergin}, \citenamefont {Pronko},\ and\ \citenamefont
  {Abarenkova}}]{izergin1998}%
  \BibitemOpen
  \bibfield  {author} {\bibinfo {author} {\bibfnamefont {A.}~\bibnamefont
  {Izergin}}, \bibinfo {author} {\bibfnamefont {A.}~\bibnamefont {Pronko}},\
  and\ \bibinfo {author} {\bibfnamefont {N.}~\bibnamefont {Abarenkova}},\
  }\bibfield  {title} {\bibinfo {title} {{Temperature correlators in the
  one-dimensional Hubbard model in the strong coupling limit}},\ }\href
  {https://www.sciencedirect.com/science/article/pii/S0375960198004423}
  {\bibfield  {journal} {\bibinfo  {journal} {Phys. Lett. A}\ }\textbf
  {\bibinfo {volume} {245}},\ \bibinfo {pages} {537} (\bibinfo {year}
  {1998})}\BibitemShut {NoStop}%
\bibitem [{\citenamefont {Abarenkova}\ and\ \citenamefont
  {Pronko}(2002)}]{abarenkova2002}%
  \BibitemOpen
  \bibfield  {author} {\bibinfo {author} {\bibfnamefont {N.~I.}\ \bibnamefont
  {Abarenkova}}\ and\ \bibinfo {author} {\bibfnamefont {A.~G.}\ \bibnamefont
  {Pronko}},\ }\bibfield  {title} {\bibinfo {title} {{Temperature correlation
  function in the absolutely anisotropic XXZ Heisenberg magnet}},\ }\href
  {https://doi.org/10.1023/A:1015480916713} {\bibfield  {journal} {\bibinfo
  {journal} {Theor. Math. Phys.}\ }\textbf {\bibinfo {volume} {131}},\ \bibinfo
  {pages} {690} (\bibinfo {year} {2002})}\BibitemShut {NoStop}%
\bibitem [{\citenamefont {Bogoliubov}\ and\ \citenamefont
  {Malyshev}(2011)}]{bogoliubov2011}%
  \BibitemOpen
  \bibfield  {author} {\bibinfo {author} {\bibfnamefont {N.~M.}\ \bibnamefont
  {Bogoliubov}}\ and\ \bibinfo {author} {\bibfnamefont {C.~L.}\ \bibnamefont
  {Malyshev}},\ }\bibfield  {title} {\bibinfo {title} {{Ising limit of a
  Heisenberg XXZ magnet and some temperature correlation functions}},\ }\href
  {https://doi.org/10.1007/s11232-011-0129-4} {\bibfield  {journal} {\bibinfo
  {journal} {Theor. Math. Phys.}\ }\textbf {\bibinfo {volume} {169}},\ \bibinfo
  {pages} {1517} (\bibinfo {year} {2011})}\BibitemShut {NoStop}%
\bibitem [{\citenamefont {Pai}\ and\ \citenamefont {Pretko}(2020)}]{pai2020}%
  \BibitemOpen
  \bibfield  {author} {\bibinfo {author} {\bibfnamefont {S.}~\bibnamefont
  {Pai}}\ and\ \bibinfo {author} {\bibfnamefont {M.}~\bibnamefont {Pretko}},\
  }\bibfield  {title} {\bibinfo {title} {Fractons from confinement in one
  dimension},\ }\href {https://doi.org/10.1103/PhysRevResearch.2.013094}
  {\bibfield  {journal} {\bibinfo  {journal} {Phys. Rev. Res.}\ }\textbf
  {\bibinfo {volume} {2}},\ \bibinfo {pages} {013094} (\bibinfo {year}
  {2020})}\BibitemShut {NoStop}%
\bibitem [{\citenamefont {Borla}\ \emph {et~al.}(2020)\citenamefont {Borla},
  \citenamefont {Verresen}, \citenamefont {Grusdt},\ and\ \citenamefont
  {Moroz}}]{borla2020}%
  \BibitemOpen
  \bibfield  {author} {\bibinfo {author} {\bibfnamefont {U.}~\bibnamefont
  {Borla}}, \bibinfo {author} {\bibfnamefont {R.}~\bibnamefont {Verresen}},
  \bibinfo {author} {\bibfnamefont {F.}~\bibnamefont {Grusdt}},\ and\ \bibinfo
  {author} {\bibfnamefont {S.}~\bibnamefont {Moroz}},\ }\bibfield  {title}
  {\bibinfo {title} {Confined phases of one-dimensional spinless fermions
  coupled to ${Z}_{2}$ gauge theory},\ }\href
  {https://doi.org/10.1103/PhysRevLett.124.120503} {\bibfield  {journal}
  {\bibinfo  {journal} {Phys. Rev. Lett.}\ }\textbf {\bibinfo {volume} {124}},\
  \bibinfo {pages} {120503} (\bibinfo {year} {2020})}\BibitemShut {NoStop}%
\bibitem [{\citenamefont {Tartaglia}\ \emph {et~al.}(2022)\citenamefont
  {Tartaglia}, \citenamefont {Calabrese},\ and\ \citenamefont
  {Bertini}}]{tartaglia2022}%
  \BibitemOpen
  \bibfield  {author} {\bibinfo {author} {\bibfnamefont {E.}~\bibnamefont
  {Tartaglia}}, \bibinfo {author} {\bibfnamefont {P.}~\bibnamefont
  {Calabrese}},\ and\ \bibinfo {author} {\bibfnamefont {B.}~\bibnamefont
  {Bertini}},\ }\bibfield  {title} {\bibinfo {title} {{Real-time evolution in
  the Hubbard model with infinite repulsion}},\ }\href
  {https://doi.org/10.21468/SciPostPhys.12.1.028} {\bibfield  {journal}
  {\bibinfo  {journal} {SciPost Phys.}\ }\textbf {\bibinfo {volume} {12}},\
  \bibinfo {pages} {028} (\bibinfo {year} {2022})}\BibitemShut {NoStop}%
\bibitem [{\citenamefont {Bastianello}\ \emph {et~al.}(2022)\citenamefont
  {Bastianello}, \citenamefont {Borla},\ and\ \citenamefont
  {Moroz}}]{bastianello2022}%
  \BibitemOpen
  \bibfield  {author} {\bibinfo {author} {\bibfnamefont {A.}~\bibnamefont
  {Bastianello}}, \bibinfo {author} {\bibfnamefont {U.}~\bibnamefont {Borla}},\
  and\ \bibinfo {author} {\bibfnamefont {S.}~\bibnamefont {Moroz}},\ }\bibfield
   {title} {\bibinfo {title} {{Fragmentation and emergent integrable transport
  in the weakly tilted Ising chain}},\ }\href
  {https://doi.org/10.1103/PhysRevLett.128.196601} {\bibfield  {journal}
  {\bibinfo  {journal} {Phys. Rev. Lett.}\ }\textbf {\bibinfo {volume} {128}},\
  \bibinfo {pages} {196601} (\bibinfo {year} {2022})}\BibitemShut {NoStop}%
\bibitem [{\citenamefont {Fagotti}(2022)}]{fagotti2022}%
  \BibitemOpen
  \bibfield  {author} {\bibinfo {author} {\bibfnamefont {M.}~\bibnamefont
  {Fagotti}},\ }\bibfield  {title} {\bibinfo {title} {Global quenches after
  localized perturbations},\ }\href
  {https://doi.org/10.1103/PhysRevLett.128.110602} {\bibfield  {journal}
  {\bibinfo  {journal} {Phys. Rev. Lett.}\ }\textbf {\bibinfo {volume} {128}},\
  \bibinfo {pages} {110602} (\bibinfo {year} {2022})}\BibitemShut {NoStop}%
\bibitem [{\citenamefont {Jones}\ and\ \citenamefont
  {Linden}(2022)}]{jones2022}%
  \BibitemOpen
  \bibfield  {author} {\bibinfo {author} {\bibfnamefont {N.~G.}\ \bibnamefont
  {Jones}}\ and\ \bibinfo {author} {\bibfnamefont {N.}~\bibnamefont {Linden}},\
  }\bibfield  {title} {\bibinfo {title} {{Integrable spin chains and the
  Clifford group}},\ }\href {https://doi.org/10.1063/5.0095870} {\bibfield
  {journal} {\bibinfo  {journal} {J. Mat. Phys.}\ }\textbf {\bibinfo {volume}
  {63}},\ \bibinfo {pages} {101901} (\bibinfo {year} {2022})}\BibitemShut
  {NoStop}%
\bibitem [{\citenamefont {Eck}\ and\ \citenamefont {Fendley}(2023)}]{eck2023}%
  \BibitemOpen
  \bibfield  {author} {\bibinfo {author} {\bibfnamefont {L.}~\bibnamefont
  {Eck}}\ and\ \bibinfo {author} {\bibfnamefont {P.}~\bibnamefont {Fendley}},\
  }\bibfield  {title} {\bibinfo {title} {{From the XXZ chain to the integrable
  Rydberg-blockade ladder via non-invertible duality defects}},\ }\href
  {https://doi.org/10.48550/arXiv.2302.14081} {\bibfield  {journal} {\bibinfo
  {journal} {arXiv:2302.14081}\ } (\bibinfo {year} {2023})}\BibitemShut
  {NoStop}%
\bibitem [{\citenamefont {Fishman}\ \emph
  {et~al.}(2022{\natexlab{a}})\citenamefont {Fishman}, \citenamefont {White},\
  and\ \citenamefont {Stoudenmire}}]{itensor}%
  \BibitemOpen
  \bibfield  {author} {\bibinfo {author} {\bibfnamefont {M.}~\bibnamefont
  {Fishman}}, \bibinfo {author} {\bibfnamefont {S.~R.}\ \bibnamefont {White}},\
  and\ \bibinfo {author} {\bibfnamefont {E.~M.}\ \bibnamefont {Stoudenmire}},\
  }\bibfield  {title} {\bibinfo {title} {{The ITensor software library for
  tensor network calculations}},\ }\href
  {https://doi.org/10.21468/SciPostPhysCodeb.4} {\bibfield  {journal} {\bibinfo
   {journal} {SciPost Phys. Codebases}\ ,\ \bibinfo {pages} {4}} (\bibinfo
  {year} {2022}{\natexlab{a}})}\BibitemShut {NoStop}%
\bibitem [{\citenamefont {Fishman}\ \emph
  {et~al.}(2022{\natexlab{b}})\citenamefont {Fishman}, \citenamefont {White},\
  and\ \citenamefont {Stoudenmire}}]{itensor-r0.3}%
  \BibitemOpen
  \bibfield  {author} {\bibinfo {author} {\bibfnamefont {M.}~\bibnamefont
  {Fishman}}, \bibinfo {author} {\bibfnamefont {S.~R.}\ \bibnamefont {White}},\
  and\ \bibinfo {author} {\bibfnamefont {E.~M.}\ \bibnamefont {Stoudenmire}},\
  }\bibfield  {title} {\bibinfo {title} {{Codebase release 0.3 for ITensor}},\
  }\href {https://doi.org/10.21468/SciPostPhysCodeb.4-r0.3} {\bibfield
  {journal} {\bibinfo  {journal} {SciPost Phys. Codebases}\ ,\ \bibinfo {pages}
  {4}} (\bibinfo {year} {2022}{\natexlab{b}})}\BibitemShut {NoStop}%
\bibitem [{\citenamefont {Sanderson}\ and\ \citenamefont
  {Curtin}(2016)}]{sanderson2016}%
  \BibitemOpen
  \bibfield  {author} {\bibinfo {author} {\bibfnamefont {C.}~\bibnamefont
  {Sanderson}}\ and\ \bibinfo {author} {\bibfnamefont {R.}~\bibnamefont
  {Curtin}},\ }\bibfield  {title} {\bibinfo {title} {Armadillo: a
  template-based c++ library for linear algebra},\ }\href
  {https://doi.org/10.21105/joss.00026} {\bibfield  {journal} {\bibinfo
  {journal} {J. Open Source Softw.}\ }\textbf {\bibinfo {volume} {1}},\
  \bibinfo {pages} {26} (\bibinfo {year} {2016})}\BibitemShut {NoStop}%
\bibitem [{\citenamefont {Sanderson}\ and\ \citenamefont
  {Curtin}(2018)}]{sanderson2018}%
  \BibitemOpen
  \bibfield  {author} {\bibinfo {author} {\bibfnamefont {C.}~\bibnamefont
  {Sanderson}}\ and\ \bibinfo {author} {\bibfnamefont {R.}~\bibnamefont
  {Curtin}},\ }\bibfield  {title} {\bibinfo {title} {A user-friendly hybrid
  sparse matrix class in c++},\ }in\ \href@noop {} {\emph {\bibinfo {booktitle}
  {Mathematical Software -- ICMS 2018}}},\ \bibinfo {editor} {edited by\
  \bibinfo {editor} {\bibfnamefont {J.~H.}\ \bibnamefont {Davenport}}, \bibinfo
  {editor} {\bibfnamefont {M.}~\bibnamefont {Kauers}}, \bibinfo {editor}
  {\bibfnamefont {G.}~\bibnamefont {Labahn}},\ and\ \bibinfo {editor}
  {\bibfnamefont {J.}~\bibnamefont {Urban}}}\ (\bibinfo  {publisher} {Springer
  International Publishing},\ \bibinfo {address} {Cham},\ \bibinfo {year}
  {2018})\ pp.\ \bibinfo {pages} {422--430}\BibitemShut {NoStop}%
\bibitem [{\citenamefont {Causer}\ \emph {et~al.}(2020)\citenamefont {Causer},
  \citenamefont {Lesanovsky}, \citenamefont {Ba\~nuls},\ and\ \citenamefont
  {Garrahan}}]{causer2020}%
  \BibitemOpen
  \bibfield  {author} {\bibinfo {author} {\bibfnamefont {L.}~\bibnamefont
  {Causer}}, \bibinfo {author} {\bibfnamefont {I.}~\bibnamefont {Lesanovsky}},
  \bibinfo {author} {\bibfnamefont {M.~C.}\ \bibnamefont {Ba\~nuls}},\ and\
  \bibinfo {author} {\bibfnamefont {J.~P.}\ \bibnamefont {Garrahan}},\
  }\bibfield  {title} {\bibinfo {title} {{Dynamics and large deviation
  transitions of the XOR-Fredrickson-Andersen kinetically constrained model}},\
  }\href {https://doi.org/10.1103/PhysRevE.102.052132} {\bibfield  {journal}
  {\bibinfo  {journal} {Phys. Rev. E}\ }\textbf {\bibinfo {volume} {102}},\
  \bibinfo {pages} {052132} (\bibinfo {year} {2020})}\BibitemShut {NoStop}%
\bibitem [{\citenamefont {Touchette}(2009)}]{touchette2009}%
  \BibitemOpen
  \bibfield  {author} {\bibinfo {author} {\bibfnamefont {H.}~\bibnamefont
  {Touchette}},\ }\bibfield  {title} {\bibinfo {title} {The large deviation
  approach to statistical mechanics},\ }\href
  {https://www.sciencedirect.com/science/article/pii/S0370157309001410}
  {\bibfield  {journal} {\bibinfo  {journal} {Phys. Rep.}\ }\textbf {\bibinfo
  {volume} {478}},\ \bibinfo {pages} {1} (\bibinfo {year} {2009})}\BibitemShut
  {NoStop}%
\bibitem [{\citenamefont {Jack}(2020)}]{jack2020ergodicity}%
  \BibitemOpen
  \bibfield  {author} {\bibinfo {author} {\bibfnamefont {R.~L.}\ \bibnamefont
  {Jack}},\ }\bibfield  {title} {\bibinfo {title} {Ergodicity and large
  deviations in physical systems with stochastic dynamics},\ }\href
  {https://doi.org/10.1140/epjb/e2020-100605-3} {\bibfield  {journal} {\bibinfo
   {journal} {Eur. Phys. J. B}\ }\textbf {\bibinfo {volume} {93}},\ \bibinfo
  {pages} {74} (\bibinfo {year} {2020})}\BibitemShut {NoStop}%
\bibitem [{\citenamefont {Lecomte}\ \emph {et~al.}(2005)\citenamefont
  {Lecomte}, \citenamefont {Appert-Rolland},\ and\ \citenamefont {van
  Wijland}}]{lecomte2005}%
  \BibitemOpen
  \bibfield  {author} {\bibinfo {author} {\bibfnamefont {V.}~\bibnamefont
  {Lecomte}}, \bibinfo {author} {\bibfnamefont {C.}~\bibnamefont
  {Appert-Rolland}},\ and\ \bibinfo {author} {\bibfnamefont {F.}~\bibnamefont
  {van Wijland}},\ }\bibfield  {title} {\bibinfo {title} {{Chaotic properties
  of systems with Markov dynamics}},\ }\href
  {https://doi.org/10.1103/PhysRevLett.95.010601} {\bibfield  {journal}
  {\bibinfo  {journal} {Phys. Rev. Lett.}\ }\textbf {\bibinfo {volume} {95}},\
  \bibinfo {pages} {010601} (\bibinfo {year} {2005})}\BibitemShut {NoStop}%
\bibitem [{\citenamefont {Lecomte}\ \emph {et~al.}(2007)\citenamefont
  {Lecomte}, \citenamefont {Appert-Rolland},\ and\ \citenamefont {van
  Wijland}}]{lecomte2007}%
  \BibitemOpen
  \bibfield  {author} {\bibinfo {author} {\bibfnamefont {V.}~\bibnamefont
  {Lecomte}}, \bibinfo {author} {\bibfnamefont {C.}~\bibnamefont
  {Appert-Rolland}},\ and\ \bibinfo {author} {\bibfnamefont {F.}~\bibnamefont
  {van Wijland}},\ }\bibfield  {title} {\bibinfo {title} {{Thermodynamic
  formalism for systems with Markov dynamics}},\ }\href
  {https://doi.org/10.1007/s10955-006-9254-0} {\bibfield  {journal} {\bibinfo
  {journal} {J. Stat. Phys.}\ }\textbf {\bibinfo {volume} {127}},\ \bibinfo
  {pages} {51} (\bibinfo {year} {2007})}\BibitemShut {NoStop}%
\bibitem [{\citenamefont {Garrahan}\ \emph {et~al.}(2009)\citenamefont
  {Garrahan}, \citenamefont {Jack}, \citenamefont {Lecomte}, \citenamefont
  {Pitard}, \citenamefont {van Duijvendijk},\ and\ \citenamefont {van
  Wijland}}]{garrahan2009}%
  \BibitemOpen
  \bibfield  {author} {\bibinfo {author} {\bibfnamefont {J.~P.}\ \bibnamefont
  {Garrahan}}, \bibinfo {author} {\bibfnamefont {R.~L.}\ \bibnamefont {Jack}},
  \bibinfo {author} {\bibfnamefont {V.}~\bibnamefont {Lecomte}}, \bibinfo
  {author} {\bibfnamefont {E.}~\bibnamefont {Pitard}}, \bibinfo {author}
  {\bibfnamefont {K.}~\bibnamefont {van Duijvendijk}},\ and\ \bibinfo {author}
  {\bibfnamefont {F.}~\bibnamefont {van Wijland}},\ }\bibfield  {title}
  {\bibinfo {title} {First-order dynamical phase transition in models of
  glasses: an approach based on ensembles of histories},\ }\href
  {https://dx.doi.org/10.1088/1751-8113/42/7/075007} {\bibfield  {journal}
  {\bibinfo  {journal} {J. Phys. A}\ }\textbf {\bibinfo {volume} {42}},\
  \bibinfo {pages} {075007} (\bibinfo {year} {2009})}\BibitemShut {NoStop}%
\bibitem [{\citenamefont {Golinelli}\ and\ \citenamefont
  {Mallick}(2006)}]{golinelli2006}%
  \BibitemOpen
  \bibfield  {author} {\bibinfo {author} {\bibfnamefont {O.}~\bibnamefont
  {Golinelli}}\ and\ \bibinfo {author} {\bibfnamefont {K.}~\bibnamefont
  {Mallick}},\ }\bibfield  {title} {\bibinfo {title} {The asymmetric simple
  exclusion process: an integrable model for non-equilibrium statistical
  mechanics},\ }\href {https://doi.org/10.1088/0305-4470/39/41/S03} {\bibfield
  {journal} {\bibinfo  {journal} {J. Phys. A}\ }\textbf {\bibinfo {volume}
  {39}},\ \bibinfo {pages} {12679} (\bibinfo {year} {2006})}\BibitemShut
  {NoStop}%
\bibitem [{\citenamefont {Appert-Rolland}\ \emph {et~al.}(2008)\citenamefont
  {Appert-Rolland}, \citenamefont {Derrida}, \citenamefont {Lecomte},\ and\
  \citenamefont {van Wijland}}]{appert-rolland2008universal}%
  \BibitemOpen
  \bibfield  {author} {\bibinfo {author} {\bibfnamefont {C.}~\bibnamefont
  {Appert-Rolland}}, \bibinfo {author} {\bibfnamefont {B.}~\bibnamefont
  {Derrida}}, \bibinfo {author} {\bibfnamefont {V.}~\bibnamefont {Lecomte}},\
  and\ \bibinfo {author} {\bibfnamefont {F.}~\bibnamefont {van Wijland}},\
  }\bibfield  {title} {\bibinfo {title} {{Universal cumulants of the current in
  diffusive systems on a ring}},\ }\href
  {https://doi.org/10.1103/PhysRevE.78.021122} {\bibfield  {journal} {\bibinfo
  {journal} {Phys. Rev. E}\ }\textbf {\bibinfo {volume} {78}},\ \bibinfo
  {pages} {021122} (\bibinfo {year} {2008})}\BibitemShut {NoStop}%
\bibitem [{\citenamefont {Jack}\ \emph {et~al.}(2015)\citenamefont {Jack},
  \citenamefont {Thompson},\ and\ \citenamefont
  {Sollich}}]{jack2015hyperuniformity}%
  \BibitemOpen
  \bibfield  {author} {\bibinfo {author} {\bibfnamefont {R.~L.}\ \bibnamefont
  {Jack}}, \bibinfo {author} {\bibfnamefont {I.~R.}\ \bibnamefont {Thompson}},\
  and\ \bibinfo {author} {\bibfnamefont {P.}~\bibnamefont {Sollich}},\
  }\bibfield  {title} {\bibinfo {title} {{Hyperuniformity and phase separation
  in biased ensembles of trajectories for diffusive systems}},\ }\href
  {https://doi.org/10.1103/PhysRevLett.114.060601} {\bibfield  {journal}
  {\bibinfo  {journal} {Phys. Rev. Lett.}\ }\textbf {\bibinfo {volume} {114}},\
  \bibinfo {pages} {060601} (\bibinfo {year} {2015})}\BibitemShut {NoStop}%
\bibitem [{\citenamefont {Lecomte}\ \emph {et~al.}(2012)\citenamefont
  {Lecomte}, \citenamefont {Garrahan},\ and\ \citenamefont {van
  Wijland}}]{lecomte2012inactive}%
  \BibitemOpen
  \bibfield  {author} {\bibinfo {author} {\bibfnamefont {V.}~\bibnamefont
  {Lecomte}}, \bibinfo {author} {\bibfnamefont {J.~P.}\ \bibnamefont
  {Garrahan}},\ and\ \bibinfo {author} {\bibfnamefont {F.}~\bibnamefont {van
  Wijland}},\ }\bibfield  {title} {\bibinfo {title} {{Inactive dynamical phase
  of a symmetric exclusion process on a ring}},\ }\href
  {http://stacks.iop.org/1751-8121/45/i=17/a=175001} {\bibfield  {journal}
  {\bibinfo  {journal} {J. Phys. A}\ }\textbf {\bibinfo {volume} {45}},\
  \bibinfo {pages} {175001} (\bibinfo {year} {2012})}\BibitemShut {NoStop}%
\bibitem [{\citenamefont {Des~Cloizeaux}\ and\ \citenamefont
  {Gaudin}(1966)}]{cloizeaux1966}%
  \BibitemOpen
  \bibfield  {author} {\bibinfo {author} {\bibfnamefont {J.}~\bibnamefont
  {Des~Cloizeaux}}\ and\ \bibinfo {author} {\bibfnamefont {M.}~\bibnamefont
  {Gaudin}},\ }\bibfield  {title} {\bibinfo {title} {Anisotropic linear
  magnetic chain},\ }\href {https://doi.org/10.1063/1.1705048} {\bibfield
  {journal} {\bibinfo  {journal} {J. Mat. Phys.}\ }\textbf {\bibinfo {volume}
  {7}},\ \bibinfo {pages} {1384} (\bibinfo {year} {1966})}\BibitemShut
  {NoStop}%
\bibitem [{\citenamefont {Yang}\ and\ \citenamefont {Yang}(1966)}]{yang1966}%
  \BibitemOpen
  \bibfield  {author} {\bibinfo {author} {\bibfnamefont {C.~N.}\ \bibnamefont
  {Yang}}\ and\ \bibinfo {author} {\bibfnamefont {C.~P.}\ \bibnamefont
  {Yang}},\ }\bibfield  {title} {\bibinfo {title} {{One-dimensional chain of
  anisotropic spin-spin interactions. II. Properties of the ground-state energy
  per lattice site for an infinite system}},\ }\href
  {https://doi.org/10.1103/PhysRev.150.327} {\bibfield  {journal} {\bibinfo
  {journal} {Phys. Rev.}\ }\textbf {\bibinfo {volume} {150}},\ \bibinfo {pages}
  {327} (\bibinfo {year} {1966})}\BibitemShut {NoStop}%
\bibitem [{\citenamefont {Koma}\ and\ \citenamefont
  {Nachtergaele}(1998)}]{koma1998}%
  \BibitemOpen
  \bibfield  {author} {\bibinfo {author} {\bibfnamefont {T.}~\bibnamefont
  {Koma}}\ and\ \bibinfo {author} {\bibfnamefont {B.}~\bibnamefont
  {Nachtergaele}},\ }\bibfield  {title} {\bibinfo {title} {The complete set of
  ground states of the ferromagnetic {XXZ} chains},\ }\href
  {https://doi.org/10.4310%2Fatmp.1998.v2.n3.a4} {\bibfield  {journal}
  {\bibinfo  {journal} {ATMP}\ }\textbf {\bibinfo {volume} {2}},\ \bibinfo
  {pages} {533} (\bibinfo {year} {1998})}\BibitemShut {NoStop}%
\bibitem [{\citenamefont {Gobert}\ \emph {et~al.}(2005)\citenamefont {Gobert},
  \citenamefont {Kollath}, \citenamefont {Schollw\"ock},\ and\ \citenamefont
  {Sch\"utz}}]{gobert2005}%
  \BibitemOpen
  \bibfield  {author} {\bibinfo {author} {\bibfnamefont {D.}~\bibnamefont
  {Gobert}}, \bibinfo {author} {\bibfnamefont {C.}~\bibnamefont {Kollath}},
  \bibinfo {author} {\bibfnamefont {U.}~\bibnamefont {Schollw\"ock}},\ and\
  \bibinfo {author} {\bibfnamefont {G.}~\bibnamefont {Sch\"utz}},\ }\bibfield
  {title} {\bibinfo {title} {Real-time dynamics in spin-$\frac{1}{2}$ chains
  with adaptive time-dependent density matrix renormalization group},\ }\href
  {https://doi.org/10.1103/PhysRevE.71.036102} {\bibfield  {journal} {\bibinfo
  {journal} {Phys. Rev. E}\ }\textbf {\bibinfo {volume} {71}},\ \bibinfo
  {pages} {036102} (\bibinfo {year} {2005})}\BibitemShut {NoStop}%
\bibitem [{\citenamefont {Mossel}\ and\ \citenamefont
  {Caux}(2010)}]{mossel2010}%
  \BibitemOpen
  \bibfield  {author} {\bibinfo {author} {\bibfnamefont {J.}~\bibnamefont
  {Mossel}}\ and\ \bibinfo {author} {\bibfnamefont {J.-S.}\ \bibnamefont
  {Caux}},\ }\bibfield  {title} {\bibinfo {title} {{Relaxation dynamics in the
  gapped XXZ spin-1/2 chain}},\ }\href
  {https://dx.doi.org/10.1088/1367-2630/12/5/055028} {\bibfield  {journal}
  {\bibinfo  {journal} {New J. Phys.}\ }\textbf {\bibinfo {volume} {12}},\
  \bibinfo {pages} {055028} (\bibinfo {year} {2010})}\BibitemShut {NoStop}%
\bibitem [{\citenamefont {Misguich}\ \emph {et~al.}(2017)\citenamefont
  {Misguich}, \citenamefont {Mallick},\ and\ \citenamefont
  {Krapivsky}}]{misguich2017}%
  \BibitemOpen
  \bibfield  {author} {\bibinfo {author} {\bibfnamefont {G.}~\bibnamefont
  {Misguich}}, \bibinfo {author} {\bibfnamefont {K.}~\bibnamefont {Mallick}},\
  and\ \bibinfo {author} {\bibfnamefont {P.~L.}\ \bibnamefont {Krapivsky}},\
  }\bibfield  {title} {\bibinfo {title} {{Dynamics of the spin-$\frac{1}{2}$
  Heisenberg chain initialized in a domain-wall state}},\ }\href
  {https://doi.org/10.1103/PhysRevB.96.195151} {\bibfield  {journal} {\bibinfo
  {journal} {Phys. Rev. B}\ }\textbf {\bibinfo {volume} {96}},\ \bibinfo
  {pages} {195151} (\bibinfo {year} {2017})}\BibitemShut {NoStop}%
\bibitem [{\citenamefont {Gamayun}\ \emph {et~al.}(2019)\citenamefont
  {Gamayun}, \citenamefont {Miao},\ and\ \citenamefont
  {Ilievski}}]{gamayun2019}%
  \BibitemOpen
  \bibfield  {author} {\bibinfo {author} {\bibfnamefont {O.}~\bibnamefont
  {Gamayun}}, \bibinfo {author} {\bibfnamefont {Y.}~\bibnamefont {Miao}},\ and\
  \bibinfo {author} {\bibfnamefont {E.}~\bibnamefont {Ilievski}},\ }\bibfield
  {title} {\bibinfo {title} {{Domain-wall dynamics in the Landau-Lifshitz
  magnet and the classical-quantum correspondence for spin transport}},\ }\href
  {https://doi.org/10.1103/PhysRevB.99.140301} {\bibfield  {journal} {\bibinfo
  {journal} {Phys. Rev. B}\ }\textbf {\bibinfo {volume} {99}},\ \bibinfo
  {pages} {140301(R)} (\bibinfo {year} {2019})}\BibitemShut {NoStop}%
\bibitem [{\citenamefont {Misguich}\ \emph {et~al.}(2019)\citenamefont
  {Misguich}, \citenamefont {Pavloff},\ and\ \citenamefont
  {Pasquier}}]{misguich2019}%
  \BibitemOpen
  \bibfield  {author} {\bibinfo {author} {\bibfnamefont {G.}~\bibnamefont
  {Misguich}}, \bibinfo {author} {\bibfnamefont {N.}~\bibnamefont {Pavloff}},\
  and\ \bibinfo {author} {\bibfnamefont {V.}~\bibnamefont {Pasquier}},\
  }\bibfield  {title} {\bibinfo {title} {{Domain wall problem in the quantum
  XXZ chain and semiclassical behavior close to the isotropic point}},\ }\href
  {https://doi.org/10.21468/SciPostPhys.7.2.025} {\bibfield  {journal}
  {\bibinfo  {journal} {SciPost Phys.}\ }\textbf {\bibinfo {volume} {7}},\
  \bibinfo {pages} {025} (\bibinfo {year} {2019})}\BibitemShut {NoStop}%
\bibitem [{\citenamefont {Sollich}\ and\ \citenamefont
  {Evans}(1999)}]{sollich1999}%
  \BibitemOpen
  \bibfield  {author} {\bibinfo {author} {\bibfnamefont {P.}~\bibnamefont
  {Sollich}}\ and\ \bibinfo {author} {\bibfnamefont {M.~R.}\ \bibnamefont
  {Evans}},\ }\bibfield  {title} {\bibinfo {title} {Glassy time-scale
  divergence and anomalous coarsening in a kinetically constrained spin
  chain},\ }\href {https://doi.org/10.1103/PhysRevLett.83.3238} {\bibfield
  {journal} {\bibinfo  {journal} {Phys. Rev. Lett.}\ }\textbf {\bibinfo
  {volume} {83}},\ \bibinfo {pages} {3238} (\bibinfo {year}
  {1999})}\BibitemShut {NoStop}%
\bibitem [{\citenamefont {Ashton}\ \emph {et~al.}(2005)\citenamefont {Ashton},
  \citenamefont {Hedges},\ and\ \citenamefont {Garrahan}}]{ashton2005}%
  \BibitemOpen
  \bibfield  {author} {\bibinfo {author} {\bibfnamefont {D.~J.}\ \bibnamefont
  {Ashton}}, \bibinfo {author} {\bibfnamefont {L.~O.}\ \bibnamefont {Hedges}},\
  and\ \bibinfo {author} {\bibfnamefont {J.~P.}\ \bibnamefont {Garrahan}},\
  }\bibfield  {title} {\bibinfo {title} {Fast simulation of facilitated spin
  models},\ }\href {https://dx.doi.org/10.1088/1742-5468/2005/12/P12010}
  {\bibfield  {journal} {\bibinfo  {journal} {J. Stat. Mech.}\ }\textbf
  {\bibinfo {volume} {2005}},\ \bibinfo {pages} {P12010} (\bibinfo {year}
  {2005})}\BibitemShut {NoStop}%
\bibitem [{\citenamefont {Keys}\ \emph {et~al.}(2011)\citenamefont {Keys},
  \citenamefont {Hedges}, \citenamefont {Garrahan}, \citenamefont {Glotzer},\
  and\ \citenamefont {Chandler}}]{keys2011}%
  \BibitemOpen
  \bibfield  {author} {\bibinfo {author} {\bibfnamefont {A.~S.}\ \bibnamefont
  {Keys}}, \bibinfo {author} {\bibfnamefont {L.~O.}\ \bibnamefont {Hedges}},
  \bibinfo {author} {\bibfnamefont {J.~P.}\ \bibnamefont {Garrahan}}, \bibinfo
  {author} {\bibfnamefont {S.~C.}\ \bibnamefont {Glotzer}},\ and\ \bibinfo
  {author} {\bibfnamefont {D.}~\bibnamefont {Chandler}},\ }\bibfield  {title}
  {\bibinfo {title} {Excitations are localized and relaxation is hierarchical
  in glass-forming liquids},\ }\href
  {https://doi.org/10.1103/PhysRevX.1.021013} {\bibfield  {journal} {\bibinfo
  {journal} {Phys. Rev. X}\ }\textbf {\bibinfo {volume} {1}},\ \bibinfo {pages}
  {021013} (\bibinfo {year} {2011})}\BibitemShut {NoStop}%
\bibitem [{Note1()}]{Note1}%
  \BibitemOpen
  \bibinfo {note} {We choose the short-time cutoff $t_{\protect \rm
  min}=10^{-2}$, but this value does not affect the estimation of the
  relaxation time.}\BibitemShut {Stop}%
\bibitem [{\citenamefont {Zadnik}\ \emph {et~al.}(2021)\citenamefont {Zadnik},
  \citenamefont {Bidzhiev},\ and\ \citenamefont {Fagotti}}]{zadnik2021hydro}%
  \BibitemOpen
  \bibfield  {author} {\bibinfo {author} {\bibfnamefont {L.}~\bibnamefont
  {Zadnik}}, \bibinfo {author} {\bibfnamefont {K.}~\bibnamefont {Bidzhiev}},\
  and\ \bibinfo {author} {\bibfnamefont {M.}~\bibnamefont {Fagotti}},\
  }\bibfield  {title} {\bibinfo {title} {{The folded spin-1/2 XXZ model: II.
  Thermodynamics and hydrodynamics with a minimal set of charges}},\ }\href
  {https://doi.org/10.21468/SciPostPhys.10.5.099} {\bibfield  {journal}
  {\bibinfo  {journal} {SciPost Phys.}\ }\textbf {\bibinfo {volume} {10}},\
  \bibinfo {pages} {099} (\bibinfo {year} {2021})}\BibitemShut {NoStop}%
\bibitem [{\citenamefont {Lesanovsky}\ and\ \citenamefont
  {Katsura}(2012)}]{lesanovsky2012}%
  \BibitemOpen
  \bibfield  {author} {\bibinfo {author} {\bibfnamefont {I.}~\bibnamefont
  {Lesanovsky}}\ and\ \bibinfo {author} {\bibfnamefont {H.}~\bibnamefont
  {Katsura}},\ }\bibfield  {title} {\bibinfo {title} {{Interacting Fibonacci
  anyons in a Rydberg gas}},\ }\href
  {https://doi.org/10.1103/PhysRevA.86.041601} {\bibfield  {journal} {\bibinfo
  {journal} {Phys. Rev. A}\ }\textbf {\bibinfo {volume} {86}},\ \bibinfo
  {pages} {041601(R)} (\bibinfo {year} {2012})}\BibitemShut {NoStop}%
\bibitem [{\citenamefont {Zadnik}\ \emph {et~al.}(2022)\citenamefont {Zadnik},
  \citenamefont {Bocini}, \citenamefont {Bidzhiev},\ and\ \citenamefont
  {Fagotti}}]{bocini2022}%
  \BibitemOpen
  \bibfield  {author} {\bibinfo {author} {\bibfnamefont {L.}~\bibnamefont
  {Zadnik}}, \bibinfo {author} {\bibfnamefont {S.}~\bibnamefont {Bocini}},
  \bibinfo {author} {\bibfnamefont {K.}~\bibnamefont {Bidzhiev}},\ and\
  \bibinfo {author} {\bibfnamefont {M.}~\bibnamefont {Fagotti}},\ }\bibfield
  {title} {\bibinfo {title} {Measurement catastrophe and ballistic spread of
  charge density with vanishing current},\ }\href
  {https://dx.doi.org/10.1088/1751-8121/aca254} {\bibfield  {journal} {\bibinfo
   {journal} {J. Phys. A}\ }\textbf {\bibinfo {volume} {55}},\ \bibinfo {pages}
  {474001} (\bibinfo {year} {2022})}\BibitemShut {NoStop}%
\bibitem [{\citenamefont {Michailidis}\ \emph {et~al.}(2018)\citenamefont
  {Michailidis}, \citenamefont {\ifmmode \check{Z}\else
  \v{Z}\fi{}nidari\ifmmode~\check{c}\else \v{c}\fi{}}, \citenamefont
  {Medvedyeva}, \citenamefont {Abanin}, \citenamefont {Prosen},\ and\
  \citenamefont {Papi\'{c}}}]{michailidis2018}%
  \BibitemOpen
  \bibfield  {author} {\bibinfo {author} {\bibfnamefont {A.~A.}\ \bibnamefont
  {Michailidis}}, \bibinfo {author} {\bibfnamefont {M.}~\bibnamefont {\ifmmode
  \check{Z}\else \v{Z}\fi{}nidari\ifmmode~\check{c}\else \v{c}\fi{}}}, \bibinfo
  {author} {\bibfnamefont {M.}~\bibnamefont {Medvedyeva}}, \bibinfo {author}
  {\bibfnamefont {D.~A.}\ \bibnamefont {Abanin}}, \bibinfo {author}
  {\bibfnamefont {T.}~\bibnamefont {Prosen}},\ and\ \bibinfo {author}
  {\bibfnamefont {Z.}~\bibnamefont {Papi\'{c}}},\ }\bibfield  {title} {\bibinfo
  {title} {Slow dynamics in translation-invariant quantum lattice models},\
  }\href {https://doi.org/10.1103/PhysRevB.97.104307} {\bibfield  {journal}
  {\bibinfo  {journal} {Phys. Rev. B}\ }\textbf {\bibinfo {volume} {97}},\
  \bibinfo {pages} {104307} (\bibinfo {year} {2018})}\BibitemShut {NoStop}%
\bibitem [{\citenamefont {Garrahan}\ and\ \citenamefont
  {Chandler}(2002)}]{garrahan2002}%
  \BibitemOpen
  \bibfield  {author} {\bibinfo {author} {\bibfnamefont {J.~P.}\ \bibnamefont
  {Garrahan}}\ and\ \bibinfo {author} {\bibfnamefont {D.}~\bibnamefont
  {Chandler}},\ }\bibfield  {title} {\bibinfo {title} {Geometrical explanation
  and scaling of dynamical heterogeneities in glass forming systems},\ }\href
  {https://doi.org/10.1103/PhysRevLett.89.035704} {\bibfield  {journal}
  {\bibinfo  {journal} {Phys. Rev. Lett.}\ }\textbf {\bibinfo {volume} {89}},\
  \bibinfo {pages} {035704} (\bibinfo {year} {2002})}\BibitemShut {NoStop}%
\bibitem [{\citenamefont {Merolle}\ \emph {et~al.}(2005)\citenamefont
  {Merolle}, \citenamefont {Garrahan},\ and\ \citenamefont
  {Chandler}}]{merolle2005}%
  \BibitemOpen
  \bibfield  {author} {\bibinfo {author} {\bibfnamefont {M.}~\bibnamefont
  {Merolle}}, \bibinfo {author} {\bibfnamefont {J.~P.}\ \bibnamefont
  {Garrahan}},\ and\ \bibinfo {author} {\bibfnamefont {D.}~\bibnamefont
  {Chandler}},\ }\bibfield  {title} {\bibinfo {title} {Space--time
  thermodynamics of the glass transition},\ }\href
  {https://www.pnas.org/doi/abs/10.1073/pnas.0504820102} {\bibfield  {journal}
  {\bibinfo  {journal} {Proc. Natl. Acad. Sci. USA}\ }\textbf {\bibinfo
  {volume} {102}},\ \bibinfo {pages} {10837} (\bibinfo {year}
  {2005})}\BibitemShut {NoStop}%
\bibitem [{\citenamefont {Berthier}\ \emph {et~al.}(2011)\citenamefont
  {Berthier}, \citenamefont {Biroli}, \citenamefont {Bouchaud}, \citenamefont
  {Cipelletti},\ and\ \citenamefont {van Saarloos}}]{berthier2011}%
  \BibitemOpen
  \bibfield  {author} {\bibinfo {author} {\bibfnamefont {L.}~\bibnamefont
  {Berthier}}, \bibinfo {author} {\bibfnamefont {G.}~\bibnamefont {Biroli}},
  \bibinfo {author} {\bibfnamefont {J.-P.}\ \bibnamefont {Bouchaud}}, \bibinfo
  {author} {\bibfnamefont {L.}~\bibnamefont {Cipelletti}},\ and\ \bibinfo
  {author} {\bibfnamefont {W.}~\bibnamefont {van Saarloos}},\ }\href@noop {}
  {\emph {\bibinfo {title} {Dynamical heterogeneities in glasses, colloids, and
  granular media}}},\ Vol.\ \bibinfo {volume} {150}\ (\bibinfo  {publisher}
  {OUP Oxford},\ \bibinfo {year} {2011})\BibitemShut {NoStop}%
\bibitem [{\citenamefont {Iadecola}\ and\ \citenamefont
  {Schecter}(2020)}]{iadecola2020}%
  \BibitemOpen
  \bibfield  {author} {\bibinfo {author} {\bibfnamefont {T.}~\bibnamefont
  {Iadecola}}\ and\ \bibinfo {author} {\bibfnamefont {M.}~\bibnamefont
  {Schecter}},\ }\bibfield  {title} {\bibinfo {title} {Quantum many-body scar
  states with emergent kinetic constraints and finite-entanglement revivals},\
  }\href {https://doi.org/10.1103/PhysRevB.101.024306} {\bibfield  {journal}
  {\bibinfo  {journal} {Phys. Rev. B}\ }\textbf {\bibinfo {volume} {101}},\
  \bibinfo {pages} {024306} (\bibinfo {year} {2020})}\BibitemShut {NoStop}%
\bibitem [{\citenamefont {Bu{\v c}a}(2023)}]{buca2023}%
  \BibitemOpen
  \bibfield  {author} {\bibinfo {author} {\bibfnamefont {B.}~\bibnamefont
  {Bu{\v c}a}},\ }\bibfield  {title} {\bibinfo {title} {Unified theory of local
  quantum many-body dynamics: Eigenoperator thermalization theorems},\ }\href
  {https://doi.org/10.48550/arXiv.2301.07091} {\bibfield  {journal} {\bibinfo
  {journal} {arXiv:2301.07091}\ } (\bibinfo {year} {2023})}\BibitemShut
  {NoStop}%
\bibitem [{\citenamefont {Borsi}\ \emph {et~al.}(2023)\citenamefont {Borsi},
  \citenamefont {Pristy{\'a}k},\ and\ \citenamefont {Pozsgay}}]{borsi2023}%
  \BibitemOpen
  \bibfield  {author} {\bibinfo {author} {\bibfnamefont {M.}~\bibnamefont
  {Borsi}}, \bibinfo {author} {\bibfnamefont {L.}~\bibnamefont
  {Pristy{\'a}k}},\ and\ \bibinfo {author} {\bibfnamefont {B.}~\bibnamefont
  {Pozsgay}},\ }\bibfield  {title} {\bibinfo {title} {Matrix product symmetries
  and breakdown of thermalization from hard rod deformations},\ }\href
  {https://doi.org/10.48550/arXiv.2302.07219} {\bibfield  {journal} {\bibinfo
  {journal} {arXiv:2302.07219}\ } (\bibinfo {year} {2023})}\BibitemShut
  {NoStop}%
\end{thebibliography}%

\end{document}